\documentclass{article}

\usepackage{arxiv}

\usepackage[utf8]{inputenc} % allow utf-8 input
\usepackage[T1]{fontenc}    % use 8-bit T1 fonts
\usepackage{hyperref}       % hyperlinks
\usepackage{url}            % simple URL typesetting
\usepackage{amssymb}
\usepackage{amsmath} 
\usepackage{booktabs}       % professional-quality tables
\usepackage{amsfonts}       % blackboard math symbols
\usepackage{nicefrac}       % compact symbols for 1/2, etc.
\usepackage{microtype}      % microtypography
\usepackage{cleveref}       % smart cross-referencing
\usepackage{lipsum}         % Can be removed after putting your text content
\usepackage{setspace}
\usepackage{graphicx}
\usepackage{natbib}
\usepackage{doi}
\usepackage{algorithm}
\usepackage{pgfplots}
\usepackage{tikz}
\usepackage{tabularx,ragged2e}
\usepackage[noend]{algpseudocode}
\usetikzlibrary{pgfplots.statistics}
\usepackage{subcaption}
\usepackage{rotating}
\newcolumntype{M}[1]{>{\arraybackslash}m{#1}}
\newcolumntype{C}[1]{>{\centering\arraybackslash}m{#1}}
\newcolumntype{N}{@{}m{0pt}@{}}

\title{Detecting Communities in Complex Networks using an Adaptive Genetic Algorithm and node similarity-based encoding}

\author{ {\hspace{1mm}Sajjad Hesamipour}\thanks{Corresponding Author} \\
	Department of Computer Engineering, Faculty of Electrical \& Computer Engineering\\
	University of Tabriz\\
	Iran \\
	\texttt{sajjadhesami95@ms.tabrizu.ac.ir} \\
	\And
	{\hspace{1mm}Mohammad Ali Balafar} \\
	Department of Computer Engineering, Faculty of Electrical \& Computer Engineering\\
	University of Tabriz\\
	Iran \\
\And
	{\hspace{1mm}Saeed Mousazadeh} \\
	Department of Computer Engineering, Faculty of Electrical \& Computer Engineering\\
	University of Tabriz\\
	Iran \\	
	
}

\renewcommand{\shorttitle}{Community detection using an adaptive genetic algorithm}

\hypersetup{
pdftitle={A template for the arxiv style},
pdfsubject={q-bio.NC, q-bio.QM},
pdfauthor={David S.~Hippocampus, Elias D.~Striatum},
pdfkeywords={First keyword, Second keyword, More},
}

\begin{document}
\maketitle

\begin{abstract}
Detecting communities in complex networks can shed light on the essential characteristics and functions of the modeled phenomena. This topic has attracted researchers of various fields from both academia and industry. Among the different methods implemented for community detection, Genetic Algorithms (GA) have become popular recently. Considering the drawbacks of the currently used \emph{locus-based} and \emph{solution-vector-based} encodings to represent the individuals, in this paper, we propose (1) a new node similarity-based encoding method to represent a network partition as an individual named \emph{MST-based}. Then, we propose (2) a new Adaptive Genetic Algorithm for Community Detection, along with (3) a new initial population generation function, and (4) a new adaptive mutation function called \emph{sine-based} mutation function. Using the proposed method, we combine \emph{similarity-based} and \emph{modularity-optimization-based} approaches to find the communities of complex networks in an evolutionary framework. Besides the fact that the proposed representation scheme can avoid meaningless mutations or disconnected communities, we show that the new initial population generation function, and the new adaptive mutation function, can improve the convergence time of the algorithm. Experiments and statistical tests verify the effectiveness of the proposed method compared with several classic and state-of-the-art algorithms.
\end{abstract}

% keywords can be removed

\section{Introduction}
\label{S:1}

Ever since the introduction of graphs by Leonhard Euler in the 18-th century, they became a revolutionary tool to model and analyze real-world phenomena. Many of the complex systems have interactional features and graphs laid the necessary ground to model them. The graphs acquired from real-world phenomena, typically have characteristics that differentiate them from random graphs. One of these characteristics is the diversity of the edge density in the different parts of the network. These densely inter-connected components could be interpreted differently based on the context of the modeled phenomena and provide valuable information about its essence. While these sub-graphs can represent individuals with common interests in social networks, they might indicate the proteins involved in a specific function on a protein interaction network. Therefore finding these groups can provide meaningful information for the experts or recommender systems to make sound decisions \cite{Gasparetti2020,ELMOUSSAOUI2019295}. The goal in community detection is to find a partition for the network that separates these densely connected parts from each other \cite{Gupta2020}. It must be noted that community detection differs from graph partitioning problem based on the predefined number of communities and their nodes. While this information is provided in the graph partitioning problems in advance, they form the major dilemmas of a community detection algorithm. Therefore, a community detection algorithm should be able to find the communities of a network in the absence of a predefined quantity of communities and their nodes \cite{PhysRevE.69.026113}. 

In recent years the popularity of community detection has increased among researchers from various fields. The need for a quality measure to evaluate the results of different methods led to a new measure called \emph{modularity}. \emph{Modularity} is a  method that evaluates the given partition for a network considering the non-randomness of intra-community edges. Despite \emph{modularity}'s popularity, there are still methods that define the quality of a given partition based on other approaches such as nodes similarity because of \emph{modularity}'s scalability problem \cite{callsificationSurvey2012,PhysRevE.84.066122}. On the other hand, there is another category of algorithms that attempts to combine different approaches.

Researchers have implemented different artificial intelligence and evolutionary algorithms (EA) for the purpose of maximization of \emph{modularity} and other measures. The GA's ability to solve various problems has brought considerable popularity for them in solving optimization problems. These methods start from random individuals, and through keeping and combining the fittest and eliminating the weak solutions, narrow the search space to desired solutions \cite{cai2016survey}. This seemingly simple logic has shown to be able to find remarkable results for complicated problems. In recent years, several EA-based methods proposed to solve community detection problems. Nearly all of the EA-based community detection algorithms can be classified on the \emph{modularity-optimization} category (In section \ref{secRW}, we discuss different methods in a detailed literature review). However, despite the extensive use of \emph{modularity}, it suffers from scalability problems, which indicates that \emph{modularity} can't detect communities smaller than a specific scale \cite{PhysRevE.84.066122}. Therefore efforts to propose another measure continue. Node \emph{similarity-based} algorithms try to solve this problem by offering alternative measures\cite{ChengZ16,HESAMIPOUR2019122354}. On the other hand, most of the GA-based community detection methods use \emph{locus-based} or \emph{solution-vector-based} representations to encode each solution, while each of them has deficiencies (we will discuss the details of the deficiencies of these methods in section \ref{Rep}).

In this paper, we introduce a new individual encoding scheme in an attempt to both overcome the deficiencies of the existing representation schemes and make the benefit of different approaches. In the new encoding scheme, called \emph{MST-based} representation, instead of using nodes and their neighbors, first, we create a weighted copy of the network using node similarity measures and form a binary chromosome by encoding its spanning tree's edges. Zero and one values of the elements of this string indicate the occurrence of a connection or detachment at the corresponding edge of the spanning tree. Then, we implement this representation to solve the community detection problem by a new adaptive genetic algorithm.

Compared with the other representations, \emph{MST-based} representation can reduce the search space by eliminating some rare possibilities and therefore directing the procedure towards much-appealing solutions. Also, considering the time limit as one of the major dilemmas of the EAs, this representation can perform faster, compared to the other methods, regarding the fact that using this representation reduces the possible values of each gene from the number of its corresponding node's neighbors to two (connected or detached edge). We will show that amount of information loss caused by this representation is negligible considering the improvements in other parameters. To further improve the convergence time of our method, we propose a new initial population generation method based on the proposed representation. The novel initial population generation function separates the network into some initial communities using a simple but effective threshold. Results confirm that this strategy can yield better initial populations and thus can reduce the convergence time of the algorithm. Finally, we introduce a new adaptive mutation function called \emph{sine-based} mutation function. The \emph{sine-based} mutation function creates the mutation probability distribution based on the distance of the edges from the borders of the community and a self-adaptive control parameter. The control parameter causes the distribution to change smoothly based on the improvement of the best individual of the population pool. Experiments approve the effectiveness of the proposed mutation function. We can summarize the contributions of this paper as follows:

\begin{itemize}
\item A new community detection method based on GA is proposed (Section \ref{ProposedAlgo});
\item We propose a new method to represent the individuals for community detection problems in GAs called \emph{MST-based} representation (Section \ref{Rep});
\item We propose a new method to generate the initial population and enhance the convergence time (Section \ref{InitPop});
\item A new adaptive mutation function called \emph{sine-based} mutation function is introduced, which adjusts probabilities based on a self-adaptive adjustment parameter (Section \ref{MUT});
\item Several experiments conducted to show the effectiveness of the the proposed method, and comparisons have been made with other methods (Section \ref{RES}).
\end{itemize}

The rest of the paper is organized as follows: in section \ref{secRW} we review the existing scientific literature of community detection subject, in section \ref{DefSec} we present some preliminaries and definitions, section \ref{ProposedAlgo} focuses on the details of the the proposed method, section \ref{RES} provides the results of our method and compares its results with other well-known algorithms. Finally, we conclude the paper in section \ref{CONC}.

\section{Related works}
\label{secRW}

Community detection methods can be categorized based on either their methodologies or their definitions from the communities \cite{callsificationSurvey2012, Gupta2020}. Souravlas et al. have also mentioned another emerging category as \emph{Data structure-based approaches} \cite{Souravlas2020}. Yet, it is possible to break each one of these categories into more specific subcategories. From the methodological point of view, community detection algorithms are mostly separated to \emph{Agglomerative} and \emph{Divisive} classes. While \emph{Agglomerative} methods start from the local structures and try to expand them to form the communities, \emph{Divisive} methods start from the full graph level and try to detect the communities by dividing the network into communities. On the other hand, because of the extreme richness of the different definitions of "community" in the literature, the definition-based approach divides the algorithms into sub-categories such as \emph{density-based}, \emph{vertex similarity-based}, \emph{action-based}, and \emph{influence propagation-based} methods. Newman and Girvan defined "community" based on the non-randomness of the edges among the nodes and proposed \emph{modularity} to measure this feature in different partitions using a null random model \cite{PhysRevE.69.026113}. While some methods try to solve the community detection problem via optimization of one of the definitions, others choose to solve the problem considering another one, hence resulting in different sub-categories. Furthermore, the recently trending \emph{Data structure-based} methods are based on the idea of transforming a network to another data structure such as trees \cite{Souravlas2020}. These methods use the new data structure either to reveal some of the hidden characteristics of the network or to reduce the complexity of the problem.

Newman and Girvan introduced \emph{modularity} along with a \emph{divisive modularity-optimization-based} community detection algorithm \cite{PhysRevE.69.026113}. In their method \cite{PhysRevE.69.026113}, they've repeatedly discarded the edges with high betweenness value from the graph and constructed a hierarchical tree, as the communities started to split. This iterative edge removal process goes on until all the edges are removed and a dendrogram formed from top to the bottom. Finally, they cut the dendrogram at a level that gives the maximum \emph{modularity}. Compared to the other measures, its lower computational complexity and higher accuracy draw the attention of the scientific society toward \emph{modularity}, which resulted in numerous algorithms based on it. Another \emph{modularity-optimization} method proposed to form dendrogram from bottom to top (\emph{agglomerative}) in \cite{PhysRevE.70.066111}. This method is well-known as Fastgreedy. It starts from the singleton communities and merges them following a greedy strategy until no other aggregation improves the \emph{modularity} further. Blondel et al. proposed another \emph{modularity-based agglomerative} method known as Louvain in \cite{Blondel_2008}. Their proposed method operates at two stages; first, it starts from the single node communities and merges them subjected to the rise in \emph{modularity}. Secondly, it merges these groups of nodes at the supernodes. Then it repeats the process from the beginning over the newly formed graph until no other aggregation could increase the \emph{modularity}. Later Traag et al. proposed an improved version of the Louvain called Leiden\cite{Traag2019}. They showed that the Louvain method can lead to disconnected communities and added a refinement stage to prevent it.

On the reverse side of the \emph{modularity-optimization} methods are the algorithms that use local structural information of the network to detect communities that do not focus on the optimization of a specific global measure. One of the widely used methods is called the Label Propagation algorithm (LPA) proposed in \cite{PhysRevE.76.036106}. LPA is one of the classic algorithms of community detection. LPA starts with assigning a label to each node of the network at the initial step, then updates their label to the most frequent label among their neighbors in a random sequence. In this algorithm, the label of each node corresponds to the community that the node belongs. This process quickly ends up in the uniquely labeled sub-graphs. A mutually beneficial domain with the community detection is link prediction. Hence, numerous \emph{similarity-based} methods use link prediction indexes to measure the similarities among the edges \cite{DAUD2020102716}. Cheng et al. show that link prediction can be used to improve the accuracy of community detection methods \cite{ChengZ16}. \cite{10.1145/2501025.2501031} describes some of these similarity measures in the network flow framework and shows that some of these similarities are related to different kinds of network flows. As a \emph{similarity-based agglomerative} method, Castrillo et al. defined a new similarity index based on the well-known \emph{cosine} similarity index \cite{castrillo2017fast}. Then, starting from the singleton communities, they merge them till no more \emph{weak} communities remain. Hesamipour et al. use one of these link prediction measures named as Adamic/Adar (AA) to determine the central nodes of the network and expand the communities around them \cite{HESAMIPOUR2019122354}. They find each central node based on its higher similarity score and its higher distance from other previously selected central nodes. After locating these nodes, they expand the communities around them by a game-theoretic \emph{agglomerative} approach. \cite{LIU2019321} also proposes a community detection method using the AA similarity measure. They use the AA to measure the similarity of nodes and form initial communities based on this similarity. Later, they merge these initial communities based on a specific attractiveness index. Some of the other methods of this category interpret the local-information in a probabilistic framework. In \cite{HAJIABADI2017188}, they use the proportion of neighbor nodes belonging to each community to estimate the dependency of each node to its own or the neighbor community and propose an overlapping community detection method. To find the communities, Infomap \cite{Rosvall1118} operates a random walk process and obtains the probabilities for a random walker to pass from each edge. Then it turns the community detection problem into a minimum length encoding problem and detects the communities using Huffman encoding. In \cite{NIKOLAEV2015154}, a method is proposed that removes the edges of the graph iteratively to find out their overall impact at the entropy of each node. They define the entropy of a specific node as the uncertainty of the destination node for a random walker if it starts from that node.

By the growing acceptance of \emph{modularity} and the methods based on its maximization, researchers turned their attention toward EAs, which were known to be proper algorithms to answer maximization problems. To the best of our knowledge, Bingol et al. made the first attempt to use EAs to detect the communities of a network \cite{tasgin2007community,tasgin2006community}. They've implemented the \emph{solution-vector} representation method to represent each individual at \cite{tasgin2006community}. A \emph{solution vector} is a $n$-dimentional vector ($n$ is the number of nodes) that stores the community identifier of each node in its corresponding element. First, they form the initial population by setting some nodes and their neighbors in the same communities. Later, they use \emph{modularity} as a fitness function and sort individuals based on their fitness values and select some individuals to merge them using the crossover function. Their one-way crossover function chooses a random community from one of the parents (source parent) and transfers it to the destination parent. Finally, they apply a uniform mutation function to change the value of a random gene. Later, the same authors proposed a measure named \emph{community variance} and employed it in a clean-up stage to promote the convergence time of their former method \cite{tasgin2007community}. As we have mentioned in Section \ref{S:1}, the goal of community detection is to find sub-graphs with dense internal connections and sparse outer links. \emph{Modularity} reflects both of these conditions inside of itself. A multi-objective EA is proposed based on these terms in \cite{SHI2012850}. Multi-objective methods attempt to optimize more than one measure simultaneously. They've decomposed \emph{modularity} to two different objectives to address both terms. Finally, they return the individual with the maximum \emph{modularity} as the solution. Ying et al. used a similar multi-objective conical area EA to detect the communities too \cite{ying2019parallel}. The similar multi-objective approach was also implemented  in \cite{GONG20124050} and \cite{LI2020100629}. Despite the fact that most of the GA-based community detection methods are categorized as \emph{modularity-optimization} methods, there are several other methods that attemp to optimize other measures. GA-net is one of the famous community detection algorithms based on evolutionary methods, that doesn’t use \emph{modularity} as its fitness function \cite{10.1007/978-3-540-87700-4_107}. This method proposes a new measure, called \emph{community score}, to quantify the quality of detected communities and implements it as a fitness function \cite{10.1007/978-3-540-87700-4_107}. Yet, the proposed measure on GA-net highly depends on its internal control parameter, and its results change dramatically depending on it. In \cite{doi:10.1177/0165551516657717}, Samie et al. propose another GA that detects communities using \emph{community score} measure (the measure proposed by Pizzuti in GA-net \cite{10.1007/978-3-540-87700-4_107}). Their method implements an additional clean-up step to absorb some \emph{weak} communities or single nodes into the others. Zarei et al. propose a heuristic method for initial population generation, an object migration automata-based method, and a hybrid algorithm based on object migration automata and GA \cite{doi:10.1111/coin.12273}. Their purpose to couple those methods is to evade from getting stuck in local optima.

On the other hand, some other methods try to reach a balance between \emph{modularity-optimization-based} and \emph{similarity-based} methods. These methods usually run in several stages and take the benefit of one of the strategies at each step. Saoud et al. \cite{SAOUD20181958} propose a method that creates initial communities using local information-based similarity indexes to minimize the uncertainty in the first stage, then in the second stage, they utilize a \emph{modularity-optimization-based} method to detect the final communities \cite{SAOUD20181958}. In \cite{li2019edmot} researchers propose a method called EdMot. Using triangle connections, EdMot forms a new graph. Then, it finds the densely connected components of this newly formed graph and adds extra edges to them. The resulted network summarizes both lower-level and higher-level connections, and it tends to highlight the communities. Another method called WATSET \cite{WATSET} proposes a meta-heuristic way that uses a fuzzy method to create a \emph{sense graph} in the first stage. In the next stage, it detects the crisp communities. CCGA \cite{said2018cc} is a GA that uses clustering coefficient to create initial population, then it uses \emph{modularity} as the fitness function and attempts to detect communities. Since some of these methods transfer the given network into a new data type using one of the measures, they can be categorized as \emph{data-structure-based} methods. In \cite{SAOUD20181958}, first, researchers propose to assign a value to each edge based on the similarity of its end nodes. After forming the weighted graph, they remove the edges with a lower weight than a specific threshold value. As a result of this process, they obtain some disconnected groups of nodes. Then in a clean-up procedure, they join smaller components to the larger ones. Next, they start to merge these larger groups based on the number of intermediate edges between them until \emph{modularity} keeps increasing. In a similar approach in \cite{SAOUD2016230}, again, they assign weights to the edges of the graph using similarity measures. Then they find the Maximum Spanning Tree (MST) of the weighted network. They exclude half of the lower weighted edges of this tree and obtain $\frac{(n-1)}{2}$ divided components. From this step on, they employ a similar approach to the previous work \cite{SAOUD20181958} to combine these components until \emph{modularity} improves. The idea of using MST was previously applied by Wu et al. \cite{WU20132265}. In their work \cite{WU20132265}, they estimate the distance of two nodes based on a heuristic approach and allocate a value to each edge of the graph. Then, they get the minimum spanning tree (mST) of the network and try to cut the tree in a point that maximizes the distance between two components. 

MSTs have been implemented in a variety of other applications and specifically in the EAs \cite{10.1007/978-3-319-54157-0_17}. Extraction of an underlying structure of the given network is a usual practice in community detection algorithms. This underlying structure is usually called \emph{skeleton-network}, and it represents a summary of the main network. Here, we treat the \emph{mST} of the network as a \emph{skeleton network}. Our method uses similarity indexes to assign weights on the edges of the network. Then it uses MSTs as the basis to introduce a new representation for the individuals at the community detection problems. Finally, using a new initial population function and a novel mutation function, it tries to detect the communities using a \emph{modularity-optimization-based} GA. Therefore, our method stands between the borderlines of \emph{similarity-based}, \emph{modularity-based}, and \emph{data-structure-based} methods. As far as we know, MSTs have not been implemented to solve community detection problems in this way before. Even though our method is designed for undirected networks but in the case of the existence of an \emph{skeleton network} (it can be any tree (cycle-less) that preserves the local characteristics of the network), it can be implemented on different networks such as \emph{temporal networks}, \emph{directed networks}, \emph{multi layer networks}, etc \cite{lee2014,sun2014,long2020}. In the following section, we provide some insights about preliminaries and definitions that are necessary to continue the discussion.

\section{Definitions}
\label{DefSec}
\emph{\textbf{Graph}}: \label{DefGraph} We use $G$ to represent a graph. A graph is nothing but a tuple of $\left(V,E\right)$, in which $V$ refers to the set of nodes, and $E$ is the set of edges that connect two nodes $E=\left[ (v,u) \vert (v,u)\in V^{2} \right]$. 
In this paper, we focus on the unweighted and undirected graphs. In these graphs, in the case of presence of an edge such as $e_{z}\in E$ between two nodes $v,u\in V$, we would have $e_{z}=(u,v)=(v,u)$.
 For each node $v$, we call the nodes that have a direct edge with $v$, set of its neighbors and show it by $\Gamma(v)$. 
 The degree of a node is equal to the number of nodes contained in the neighborhood of that node ($d_{v}=\vert\Gamma(v)\vert$). 
 Also, we use $\vert V \vert=n$ and $\vert E \vert=m$ to represent the number of nodes and edges of a graph.
The same definition differs slightly in the weighted networks. A weighted network is a triplet such as $G_{w}=(V,E,w)$ consisting of a set of nodes ($V$), a set of edges ($E$), and a mapping function ($w$) that maps each edge to a real number ($w:E\rightarrow \mathbb{R}$).

\emph{\textbf{Community}}: \label{DefCommunity} Although there are different definitions for the notion of the community \cite{HESAMIPOUR2019122354,FORTUNATO201075}, yet there seems to be a consensus on the definition of a community as the nodes that are densely interconnected and have fewer outer links. Therefore, assuming $C_{i}$ as the $i$-th community of a network, the following should apply to it:

\begin{equation}
\begin{split}
\sum_{i=1}^{k}{\vert \left[ (u,v)\vert(u,v)\in E , u \in C_{i}, v \in C_{i} \right] \vert} \gg \\ \sum_{i,j}{\vert \lbrace (u,v) \vert (u,v)\in E , u \in C_{i}, v \in C_{j} , i \neq j \rbrace \vert}
\end{split}
\end{equation}
In the cases that a node could become a member of more than one community, we would call them overlapping communities. In this paper, we concentrate on non-overlapping communities ($C_{i} \cap C_{j} = \varnothing$). Additionally, having that a node should be a member of at least one community, we have $ \bigcup_{i=1}^{k} C_{i} = V$. The set of non-overlapping communities that divide a network to subgraphs is called a partition $P=\lbrace C_{1},C_{2},…,C_{k} \rbrace ; i \neq j$ for $ i,j \in \{1,\dots,k\}$.

\emph{\textbf{Minimum/Maximum Spanning Trees}}: \label{DefMST} A minimum/maximum spanning tree is a tree derived from a weighted graph that involves all of the graph’s nodes and a subset of its edges that the sum of their weights is the minimum/maximum possible value. %In the graphs that each edge has a unique weight, the resulting spanning tree is also unique.%
A spanning tree always has $\vert E \vert=n-1$ edges where cutting each of them divides the tree. In the proposed method, we take the benefit of this feature, and we use the terms of \emph{Broken edge} or \emph{Border Edge} to denote the edges that were cut to split the spanning tree.  In this paper, we denote a Maximum Spanning Tree with \emph{MST} and a Minimum Spanning Tree with \emph{mST}. Naturally, one can construct both of them with the same method by simply inverting the weights of edges. Here, we use the well-known \emph{Prim} algorithm to find the corresponding \emph{MST} of a graph \cite{prim1957}.

\begin{table}%[cols=3,pos=h]%,cols=3,pos=h
\caption{Some of the widely used node similarity measures.}\label{tbl1}
\begin{tabular*}{5.8in}{p{1.2in} p{2.2in} p{2in}}%%{C C p{6cm}} 
\hline
Measure & Formula & Description\\
\hline
Common Neighbors & $\vert\Gamma_{v} \cap \Gamma_{u}\vert$ & A simple method that calculates the similarity of two nodes based on the number of their common neighbors \cite{lu2011link}.  \\
Jaccard & \begin{large}
$\frac{\vert\Gamma_{v} \cap \Gamma_{u}\vert}{\vert\Gamma_{v} \cup \Gamma_{u}\vert}$ 
\end{large}
& This method brings the similarity score of two nodes between 0 and 1 by dividing the number of common neighbors among them to the number of nodes in the union of their neighborhoods \cite{jaccard1901etude}.  \\
Cosine similarity & \begin{large} $\frac{\vert
\Gamma_{v} \cap \Gamma_{u}\vert}{\sqrt{\vert\Gamma_{v}\vert \times \vert\Gamma_{u}\vert}}$ \end{large}& This measure computes the cosine of the degree between two columns of the adjacency matrix \cite{salton1983introduction}.  \\
Hub Promoted Index (HPI) & \begin{large} $\frac{\vert\Gamma_{v} \cap \Gamma_{u}\vert}{\min{ \{\vert\Gamma_{v}\vert} ,{ \vert\Gamma_{u}\vert \} } }$ \end{large} & This measure is designed to maximize the impact of higher degree nodes \cite{ravasz2002hierarchical}.  \\
Adamic/Adar (AA) & \begin{large}$\sum_{k \in \Gamma_{v} \cap \Gamma_{u}}{\frac{1}{\vert\log{\Gamma_{k}}\vert}}$ \end{large}& This measure reduces the impact of very high degree nodes by summing the inverse logarithm of the degrees of nodes \cite{adamic2003friends}. \\
Resource Allocation Index (RA) &  \begin{large}$\sum_{k \in \Gamma_{v} \cap \Gamma_{u}}{\frac{1}{\vert{\Gamma_{k}}\vert}}$\end{large} & This measure calculates the possibility that the intermediate nodes pass the resource they have taken from u to v \cite{zhou2009predicting}. \\
Common Neighbors Degree Penalization (CNDP) &  \begin{large}$\sum_{k \in \Gamma_{v} \cap \Gamma_{u}}{\vert{C}_{k}\vert{\vert{\Gamma_{k}}\vert}^{-\beta \overline{C}}}$\end{large} & This method computes the similarity between two nodes using the common neighbors of $k$ which consist of the common neighbors of $u$ and $v$ ($\vert{C}_{k}\vert$), number of neighbors of $k$ ($\vert\Gamma_{k}\vert$) with the power of average clustering coefficient ($\overline{C}$) times a constant($\beta$, we defined the $\beta=1.76$) \cite{RAFIEE2020122950}. \\
Similarity based on Random Walk (SRW) &  \begin{large}$\sum_{l=2}^{T}{\frac{\vert\Gamma_u\vert}{2\vert E\vert}}\cdot \pi_{u,v}(l)+{\frac{\vert\Gamma_v\vert}{2\vert E\vert}}\cdot \pi_{v,u}(l)$\end{large} & Here, the random walk transition matrix of the graph ($\pi$)is used to compute the similarity between two nodes. We have set $T=5$ in our experiments\cite{Liu_2010}. \\
Hybrid Influence of Neighbors (HIN) &  \begin{large}$\sum_{l=2}^{T}{\frac{\sqrt{\overline{{\vert\Gamma\vert}_u} \cdot \overline{h_u}}}{2\vert E\vert}}\cdot \pi_{u,v}(l)+{\frac{\sqrt{\overline{{\vert\Gamma\vert}_v} \cdot \overline{h_v}}}{2\vert E\vert}}\cdot \pi_{v,u}(l)$\end{large} & This measure is a generalization of the SRW measure which uses both the average degree and the average h-indexs of the neighbors. We have set $T=5$ in our experiments\cite{HIN2020}. \\
\end{tabular*}

\end{table} 

\section{Proposed Algorithm}
\label{ProposedAlgo}
In this section, we illustrate the details of the proposed method.  First, we start by explaining the precomputation and encoding procedure in section \ref{Rep}. Next, we describe the initial population generation method in \ref{InitPop}, then we define the details of selection, crossover, and mutation functions in section \ref{SELECT} to \ref{MUT}, respectively.
\subsection{Representation}
\label{Rep}

\begin{figure*}[t!]
	\centerline{
		\includegraphics[width=6in,height=4.5in]{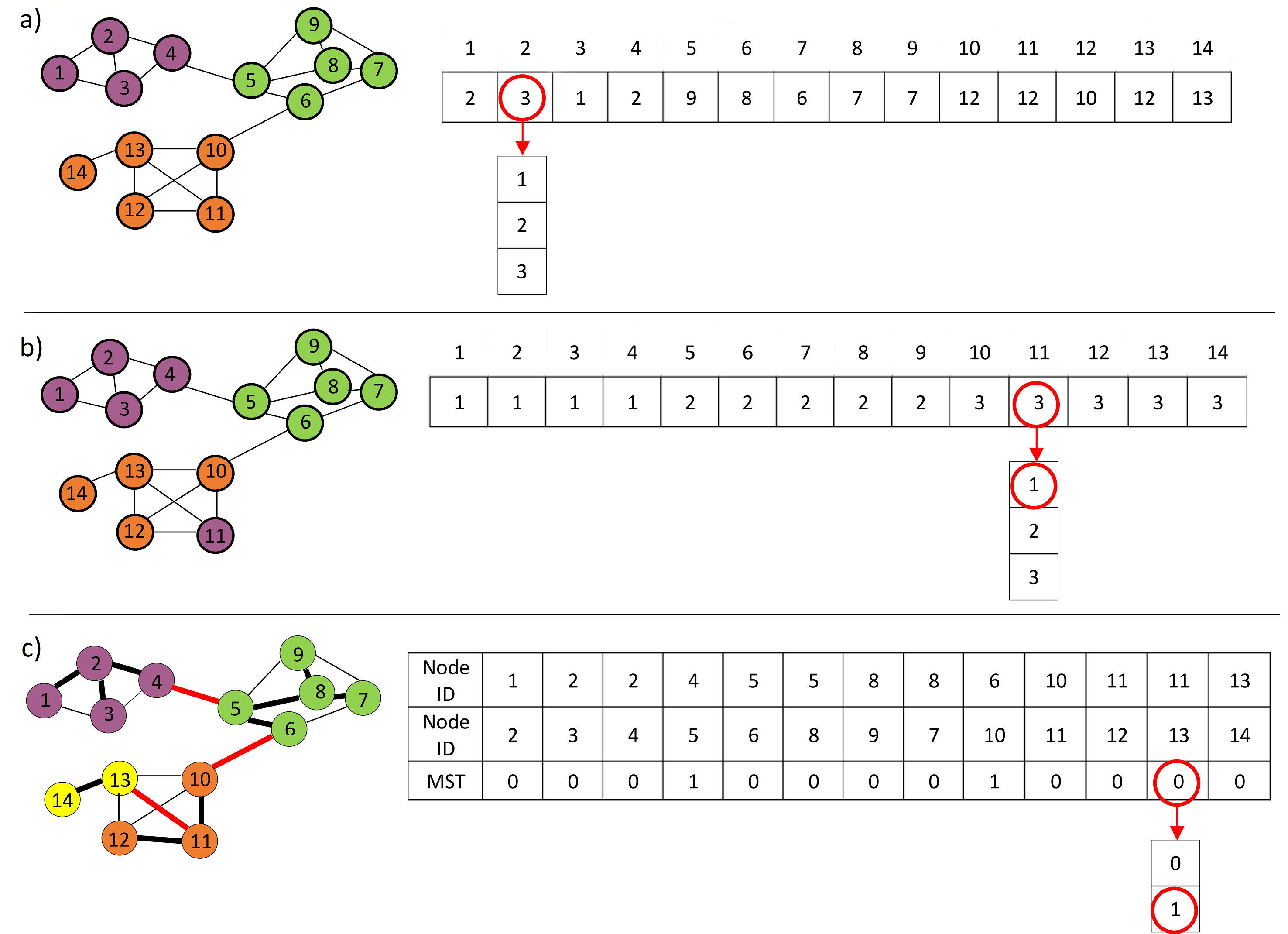}}
	\caption{Comparison of different representation schemes and their corresponding graph; a) An individual created by the locus-based method: As it could be seen using this method might cause meaningless mutations, b) An individual created by the solution-vector method: Using this method might result into separated communities, c) MST-based representation: This method reduces the domain of each element to binary and avoids meaningless mutations and separated communities.}
	\label{FIG:1}
\end{figure*}

Most of the GA-based methods in community detection, use \emph{locus-based} representation to encode the individuals of the population \cite{GONG20124050}. In this representation scheme, each individual consists of $n$ genes, each one referring to one of the nodes of the network. In the \emph{locus-based} method, the value of each gene ($g_{v}=u$) refers to the identifier of one of the nodes and specifies that these two nodes belong to the same community ($u,v \in C_{i}$). In the practical implementations, often they change the domain of each gene to the neighborhood of the corresponding node ($g_{v} \in \Gamma_{v}$). Although using a \emph{locus-based} representation may be necessary for other applications, but even this domain can have redundancies for a community detection problem. For Example, in Figure-\ref{FIG:1}.a, selecting different neighbors for node \emph{\#2} would not affect the outcome, because all of its neighbors belong to the same community. Yet, if a mutation happens on this node, the fitness value of the individual should be recomputed. Besides, if we use the \emph{locus-based} representation, we should make an additional random/educated choice from the scope of the neighbors of each node (gene). Another method that is used far less than \emph{locus-based} representation is \emph{solution-vector} representation. This method, again, holds a gene for each node in every individual. But, in this scheme holding that the graph has $k$ communities, one has to choose a value between $1$ and $k$ for each gene, indicating the identifier of the community that the node belongs to \cite{doi:10.1111/coin.12273,SHI2012850}. Although this method eliminates the need to label the members of a community in an additional pass, it might result in disconnected communities and violate the definition of community (\ref{DefCommunity}).  Figure-\ref{FIG:1}.b shows such a scenario.

To avoid such errors, we propose a new method to represent individuals. In this method, which we call it \emph{MST-Based} representation, with the cost of sacrificing some possible combinations of the nodes in a community, some problems such as effectless mutations or disconnected communities get eliminated. Furthermore, by reducing the domain of individual representation to binary, our method brings community detection problems closer to the classic GA representations \cite{sampson1976adaptation}. To create such a representation first, we assign a value for each edge of the graph using one of the node similarity/link prediction measures. Table-\ref{tbl1} lists some of the well-known and recent measures. In the resulted weighted graph, the weight of each edge denotes the similarity of two end nodes of that edge. Each of these similarity measures considers a specific feature, and therefore using a different measure might affect the results because each measure focuses on the different characteristics. After assigning weights to the edges of the graph, we choose a corresponding MST of its (practically most of the networks satisfy the unique MST criteria). As we have described in section \ref{DefCommunity}, each MST has $n-1$ edges, where discarding each of them splits the tree into two parts. Instead of encoding individuals based on the nodes, we encode them based on the edges. Hence in \emph{MST-based} representation, each individual will have $n-1$ genes with a binary domain. The value of each gene expresses the state of the corresponding edge on the MST tree. We use $1$ to indicate a broken edge and $0$ to refer to its connectedness. The \emph{MST-based} representation of a given partition for the example graph is shown in the Figure-\ref{FIG:1}-c. As described in the figure, using \emph{MST-based} representation causes a significant change by each mutation, where even a mutation over an internal edge such as $(13, 14)$, results in a brand new partition. It is obvious that using such a representation makes some combinations impossible. For example, because of the absence of an edge between node $\#5$ and $\#9$ in the resulted MST, reaching a community consisting of just these two nodes would be impossible. But in practice, considering the mutually beneficial relationship between the community detection and the node similarity/link prediction, having communities with less similar nodes is unlikely to be desirable for us \cite{DAUD2020102716, ChengZ16}. Therefore, considering the fact that the weighting process has already highlighted the edges with more similar neighbors, we can be assured that we are only neglecting the communities that might have sparse intra-community connections. Consequently, \emph{MST-based} representation can reduce the search scope of the problem. Thus the amount of information loss of using this representation is acceptable regarding its various advantages compared to the other representation schemes. On the other hand, this representation can be applied to different types of networks (such as \emph{temporal networks}, \emph{directed networks}, etc.) using an application-specific \emph{skeleton networks} instead of \emph{mST}.

\subsection{Initial population generation}
\label{InitPop}
\begin{algorithm}[t!]
    \caption{The pseudo-code of initial population generation explained in \ref{InitPop}}
    \label{algo1}
    \hspace*{\algorithmicindent} \textbf{Input:} \emph{$MST$ of the network} \\
 \hspace*{\algorithmicindent} \textbf{Output:} \emph{returns a vector representing an individual of initial population ($\overrightarrow{g}$)}
\begin{algorithmic}[1] % The number tells where the line numbering should start
\Procedure{InitGen}{$MST$}  
%\State $\overrightarrow{R}\gets $ a random sequence of $MST$ edges
%\State $\overrightarrow{E}\gets $ a zero vector with the length of $n-1$ \Comment{indicating if an edge has been expanded or not}
\State $\overrightarrow{V}\gets $ 	a zero vector with the length of $n$  \Comment{indicating if a node has been visited or not}
\State $\overrightarrow{g}\gets $ a vector that represents genes of an individual \Comment{$n-1$ genes for $n-1$ edges in the MST}
\State $B \gets $ add a random edge to the queue
\While{$B$ isn't empty}
\State $edge \gets B.pop()$
\State $node1 \gets edge.end\_node[0]$
\State $node2 \gets edge.end\_node[1]$
\If{$\overrightarrow{V}_{node1}=0$}
\State $(b,\overrightarrow{V},\overrightarrow{g}) \gets BFS\_mod(node1,node2,\overrightarrow{V},\overrightarrow{g})$
\State {add the new broken edges ($b$) to $B$}
\ElsIf{$\overrightarrow{V}_{node2}=0$}
\State $(b,\overrightarrow{V},\overrightarrow{g}) \gets BFS\_mod(node2,node1,\overrightarrow{V},\overrightarrow{g})$
\State {add the new broken edges ($b$) to $B$}
\EndIf
\EndWhile       
\EndProcedure
\end{algorithmic}
\end{algorithm}
\begin{algorithm}[t!]
    \caption{The pseudo-code of the modified version of the BFS algorithm to produce initial population generation explained in \ref{InitPop}}
    \label{algo2}
    \hspace*{\algorithmicindent} \textbf{Input:} $sNode$ : the direction that we will start to expand the edges;\\
\hspace*{\algorithmicindent}\hspace*{\algorithmicindent}
\hspace*{\algorithmicindent} $nNode$: the neighbor node;\\
\hspace*{\algorithmicindent}\hspace*{\algorithmicindent}
\hspace*{\algorithmicindent} $\overrightarrow{V}$: the \emph{visited} vector of the nodes;\\
\hspace*{\algorithmicindent}\hspace*{\algorithmicindent}
\hspace*{\algorithmicindent} $\overrightarrow{g}$: the \emph{genes} vector.\\
 \hspace*{\algorithmicindent} \textbf{Output:} \emph{returns the new broken edges($b$), updated visited vector($\overrightarrow{V}$) and the updated genes vector($\overrightarrow{g}$)}
 \begin{algorithmic}[1] % The number tells where the line numbering should start

\Procedure{BFS-mod}{$sNode$,$nNode$, $V$,$g$}  
%\State $\overrightarrow{R}\gets $ a random sequence of $MST$ edges
%\State $\overrightarrow{E}\gets $ a zero vector with the length of $n-1$ \Comment{indicating if an edge has been expanded or not}
\State $\overrightarrow{S}\gets sNode$ 	
\State $\overrightarrow{N}\gets nNode$
\State $t=0$
\While{$\overrightarrow{S}$ isn't empty}
\State $n1 \gets \overrightarrow{S}.pop()$
\State $n2 \gets \overrightarrow{N}.pop()$
\If{$t < \sqrt{n}$}
\State $\overrightarrow{V}_{n1}=1$
\State $t=t+1$
\For{each $node \in$ $\Gamma_{n1}$}
\If{$\overrightarrow{V}_{node}$}
\State add $node$ to $\overrightarrow{S}$
\State add $n1$ to $\overrightarrow{N}$
\Else
\If{$\overrightarrow{g}_{n1,node}=1$}
\State{add $(n1,node)$ to $b$}
\EndIf
\EndIf
\EndFor
\Else
\State{$\overrightarrow{g}_{n1,node} \gets 1$}
\State{add $(n1,node)$ to $b$}
\EndIf
\If{$t < \sqrt{n}$}
\State{$randEdge \gets$ a random edge from $b$}
\State{$\overrightarrow{g}_{randEdge} \gets 0$}
\EndIf
\EndWhile
\State{return $(b,\overrightarrow{V},\overrightarrow{g})$}
\EndProcedure

\end{algorithmic}
\end{algorithm}

\begin{figure*}[t!]
\begin{subfigure}[t]{1.2in}
  \includegraphics[width=1.2in,height=1.6in]{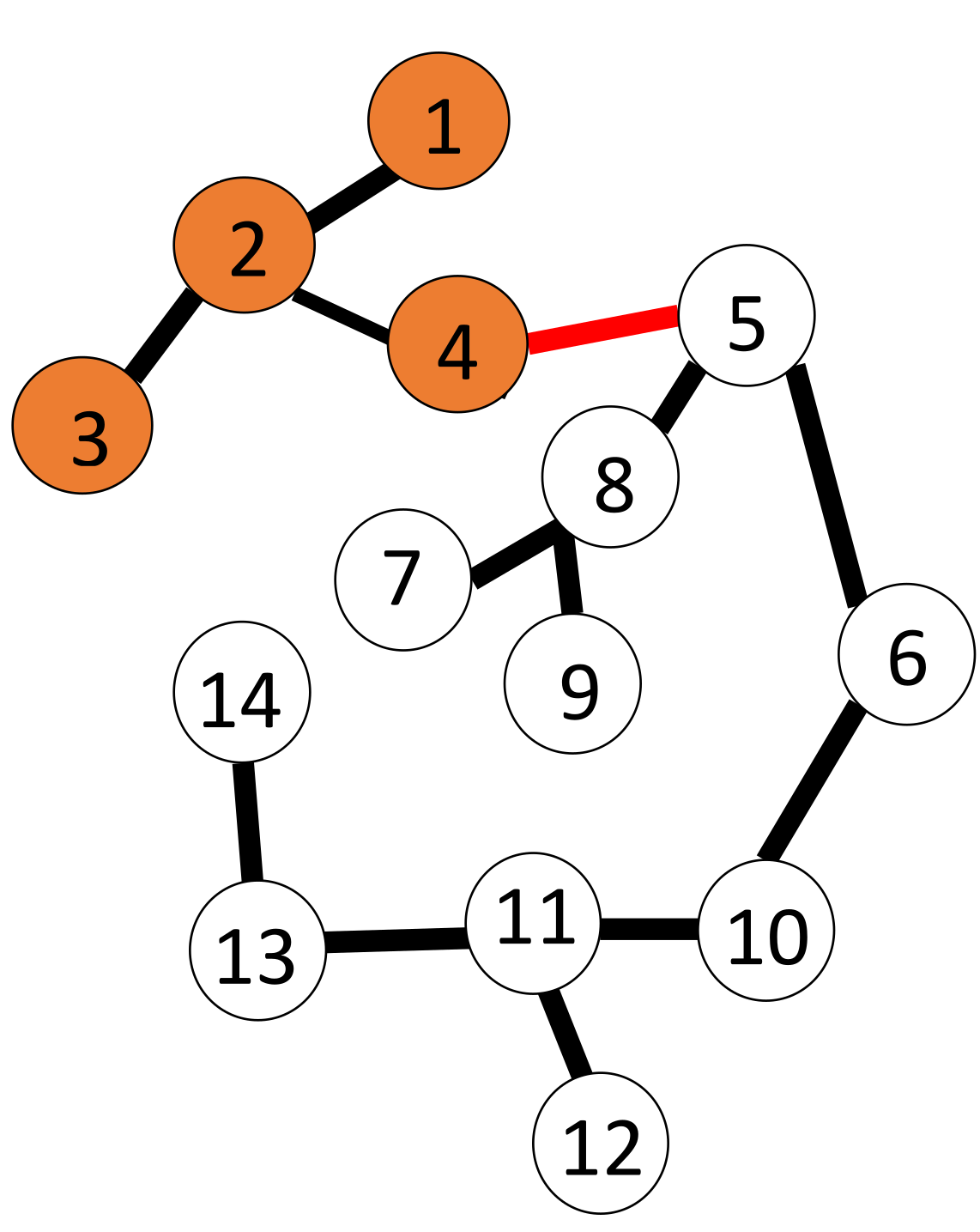}
  \caption{}
  \label{fig2:sfig1}
\end{subfigure}%
\begin{subfigure}[t]{1.2in}
  \includegraphics[width=1.2in,height=1.6in]{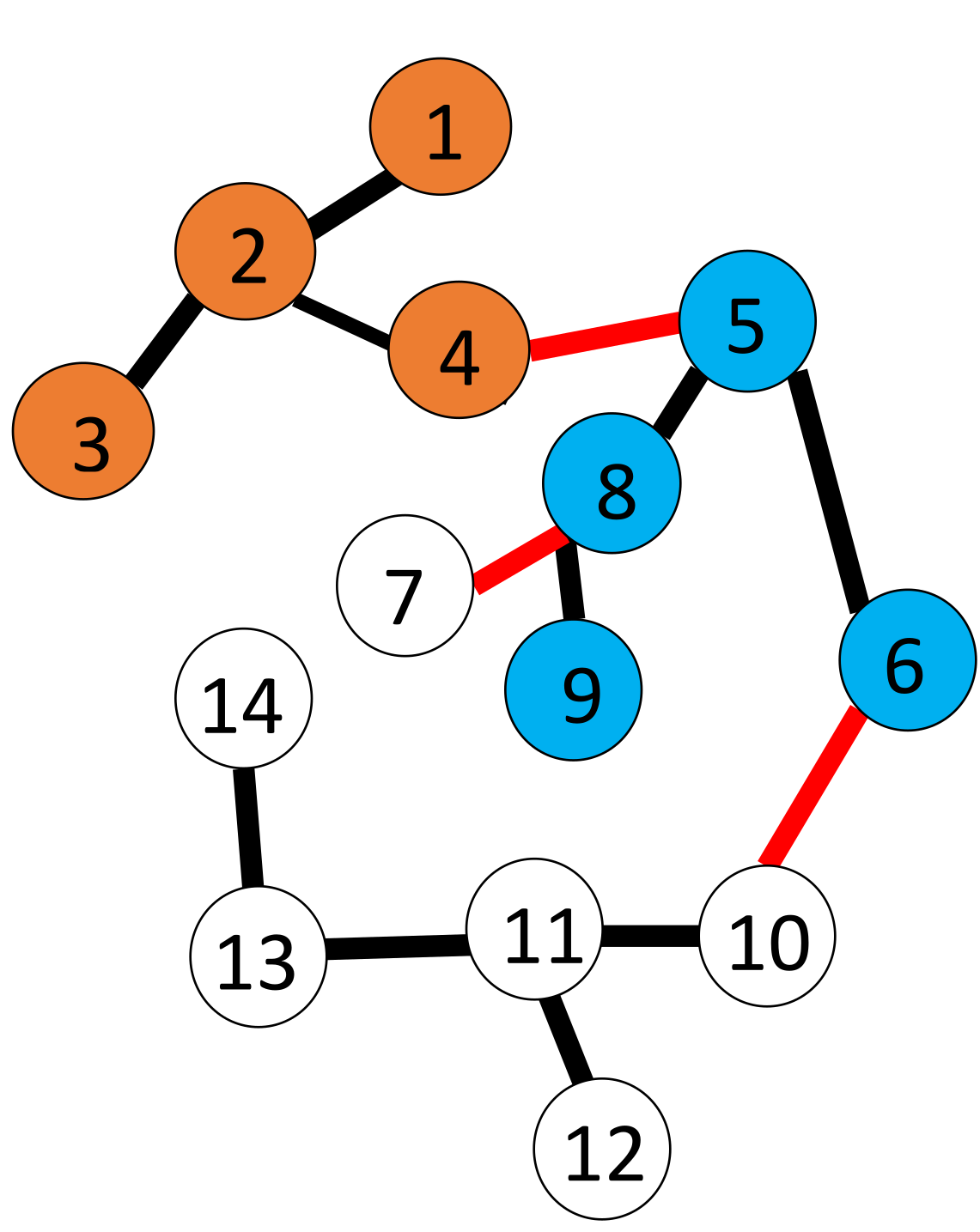}
  \caption{}
  \label{fig2:sfig2}
\end{subfigure}%
\begin{subfigure}[t]{1.2in}
  \includegraphics[width=1.2in,height=1.6in]{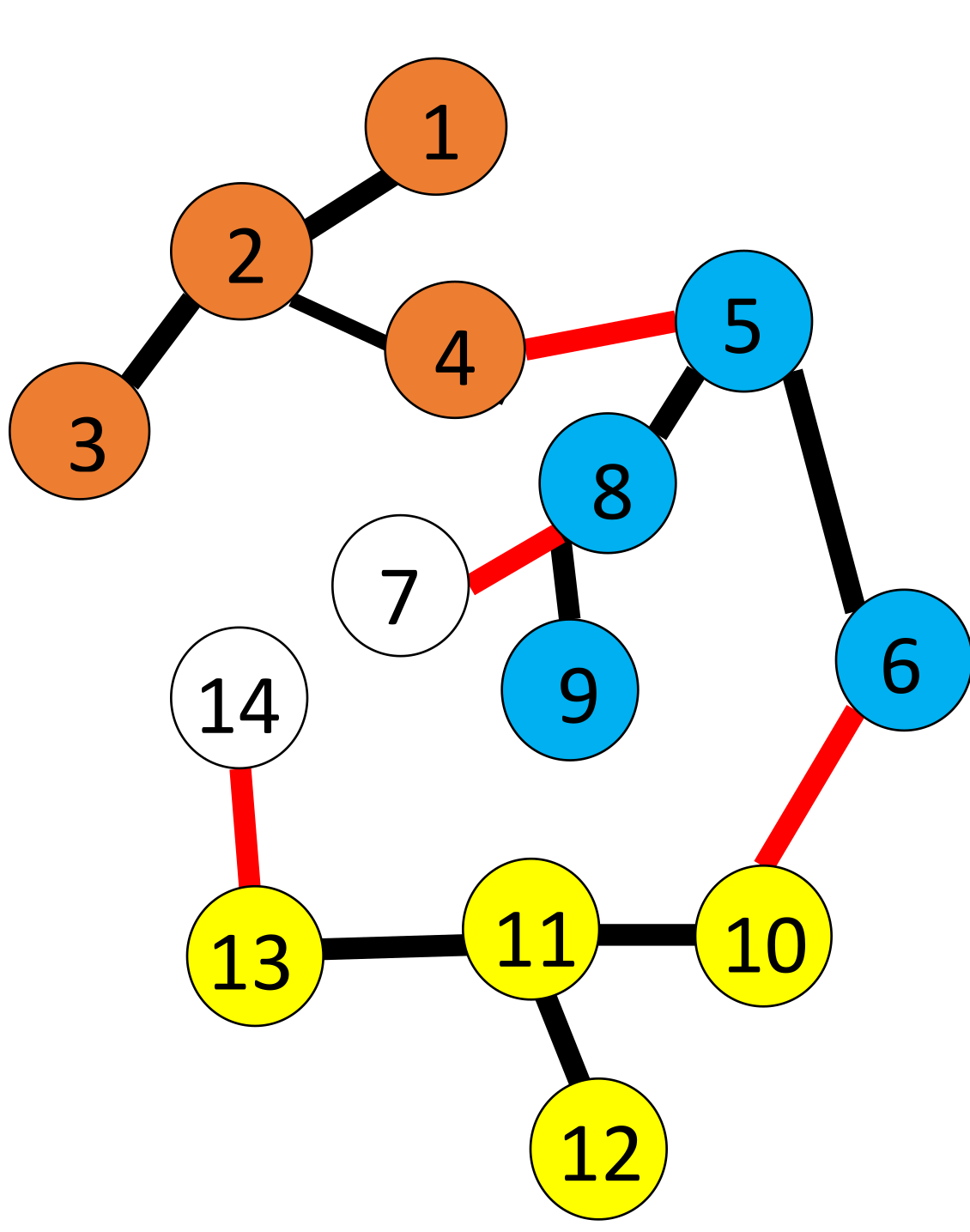}
  \caption{}
  \label{fig2:sfig3}
\end{subfigure}%
\begin{subfigure}[t]{1.2in}
  \includegraphics[width=1.2in,height=1.6in]{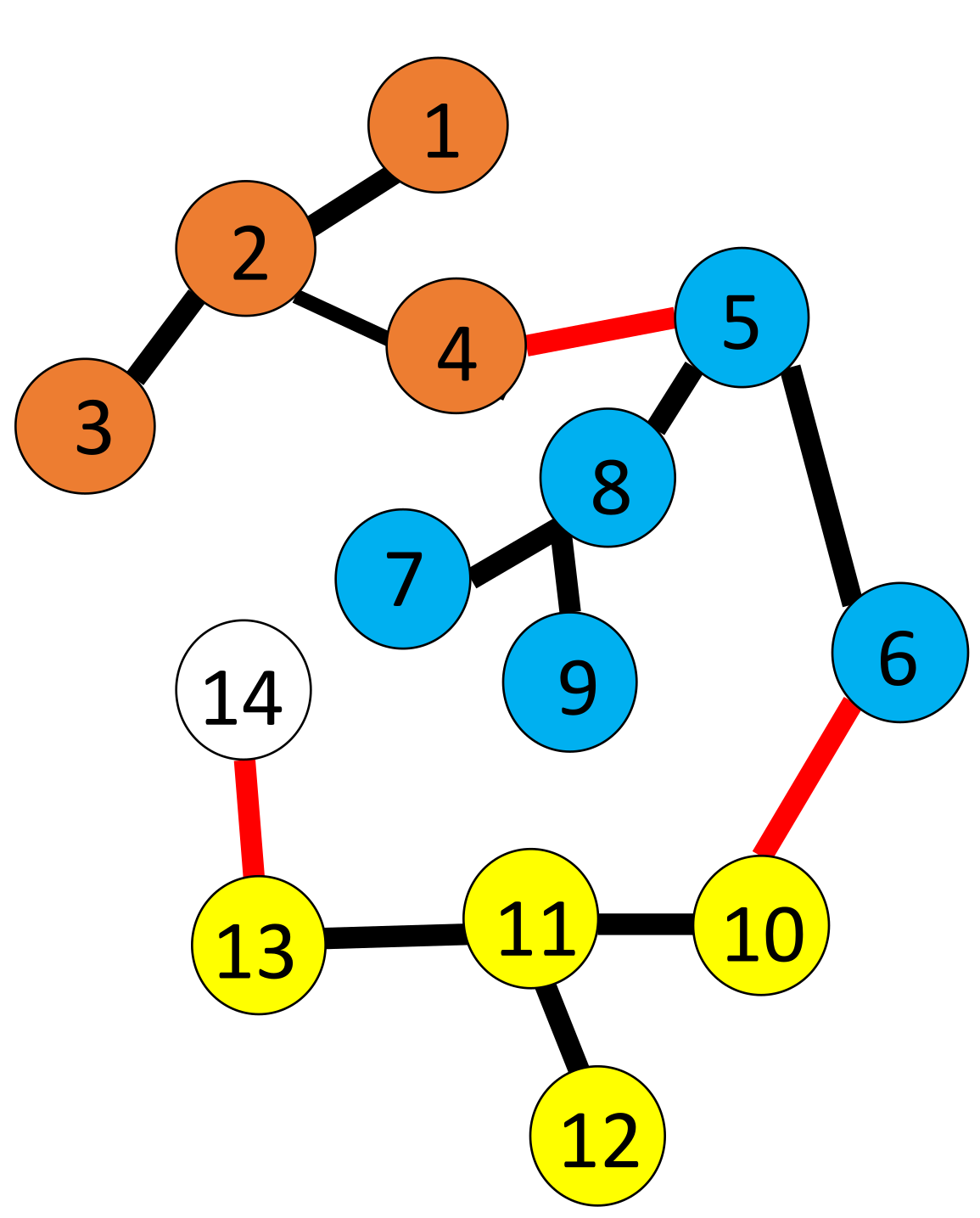}
  \caption{}
  \label{fig2:sfig4}
\end{subfigure}%
\begin{subfigure}[t]{1.2in}
  \includegraphics[width=1.2in,height=1.6in]{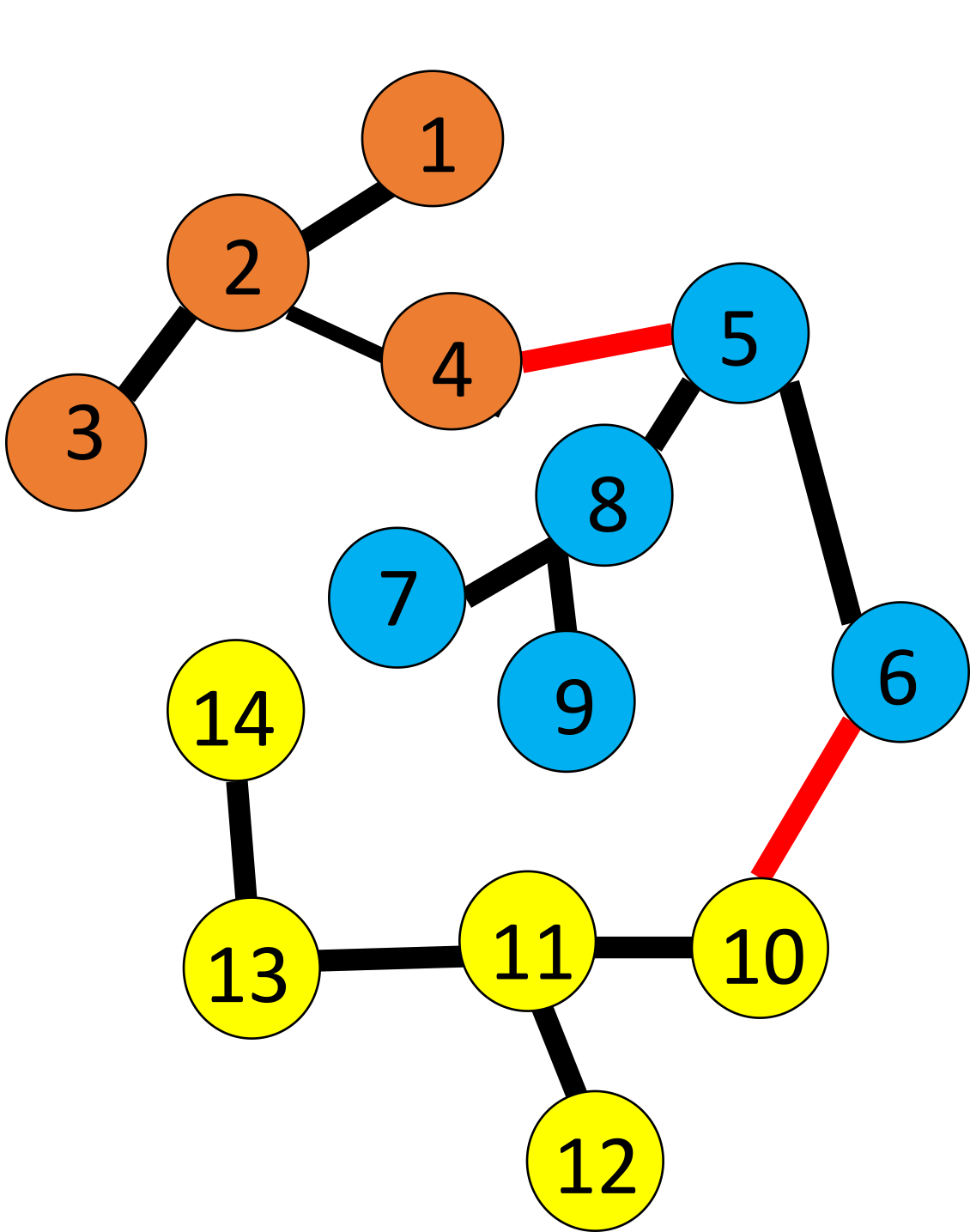}
  \caption{}
  \label{fig2:sfig5}
\end{subfigure}%
	\caption{The initial population generation procedure is explained from left to right; a)The edge ($4$, $2$) gets chosen, and its end-nodes get assigned to the corresponding community. The edges of node $2$ get passed, and, because 4 nodes are assigned to the community, procedure reaches  the predefined threshold ($\left\lceil \sqrt{14} \right\rceil=4$), and chooses another edge; b) The edge ($5$, $8$) gets chosen, and its end-nodes get assigned to the community. The broken edge from $5$ to $4$ gets ignored, but node $6$ gets added to the community. The ($8$,$9$) edge gets processed, and by adding the node $9$, community members reach to the limit and the rest of the edges in the queue get broken; c) ($10$,$11$) edge gets chosen, and the community expands around the node $11$; d) The edge between $7$ and $8$ gets chosen, and because it doesn’t reach the threshold limit, the broken edge gets connected again; e) ($13$,$14$) gets chosen, and because the community members fail to reach the threshold value, the broken edge gets reconnected.}
	\label{FIG:2}
\end{figure*}

Although GAs usually perform the search process for finding an optimal (or at least a near-optimal) solution of the problem in a manner of eliminating the weakest and survival of the fittest individual of a random pool, if we could start the process from a more rational initial population, we could improve the convergence time \cite{tasgin2007community,tasgin2006community}. Even though the number of nodes of each community depends on the edges among them, observations suggest that the size of communities could increase depending on the size of the network. Considering the cluster/community size equal to $\sqrt{n}$ ($n$ is the number of nodes) is a common hypothesis in clustering and community detection algorithms \cite{678624,BILAL201789}. We take this assumption to form initial individuals. First, we define a \emph{visited} flag for each node. Then we start by selecting a random edge. If the end nodes of the selected edge were not \emph{visited} before, then we explore the other edges connected to its end-nodes, while, keeping track of the number of nodes we have discovered so far. We keep exploring the edges in a Breadth-First Search (BFS) order until either a broken edge reached (an edge which its corresponding gene has a value of 1) or the number of the nodes of the community exceeds the threshold of $\sqrt{n}$. After exploring each edge, we set its end nodes \emph{visited} flag as checked and consider both of them as members of the same community. After reaching the determined threshold, we set the value of corresponding genes of all the remaining edges in the queue equal to $1$, meaning that these are \emph{broken edges}. In the case that the process reaches to the broken edges before the number nodes reach $\sqrt{n}$, and it can’t find a non-broken edge to expand the community, then we change the gene value of one of the broken edges to $0$ (meaning that we connect it again). We repeat this procedure until all of the MST edges are processed. Figure-\ref{FIG:2} shows the process of forming an initial individual for the example network of Figure-\ref{FIG:1}. The random selection order of edges to create such an individual is \emph{[(4,2), (5,8), (10,11), (9,8), (3,2), (7,8), (11,13), (12,11), (1,2), (6,10), (14,13), (5,4), (5,6)]}. The pseudocode of creating an individual for the initial population is given in Algorithm-\ref{algo1} and Algorithm-\ref{algo2} . The \emph{BFS-mod} function in Algorithm-\ref{algo2} is a slightly modified version of the \emph{BFS} algorithm that stops the procedure and returns the \emph{broken edges}, the updated \emph{visited} vector, and the updated chromosome after reaching the termination conditions. Experimental results prove that individuals created using this method have better fitness values, and therefore, can accelerate the convergence (section \ref{COMP}). 

\subsection{Selection}
\label{SELECT}

The goal of the selection phase is to keep the appropriate individuals and discard the weak ones. A fitness function is necessary to evaluate the quality of individuals. \emph{Modularity} is one of the most used measures to determine the quality of a partition on a graph. Most of the GAs use this measure as their fitness function for community detection problems. The details about the \emph{modularity} are described in section \ref{MEAS}. One of the important characteristics of \emph{modularity} is its additiveness \cite{FORTUNATO201075}. An additive function could be written as the sum of another elementary function. In other words, for a partition such as $P$, an additive measure ($Q$) has an elementary function like $q(.)$ such that:
\begin{singlespace}
\begin{equation}
Q(P)= \sum_{C \in P}{ q\left( C \right)}.
\label{EQ2}
\end{equation}
\end{singlespace}
The $q\left( C \right)$ function for modularity is like the following:
\begin{equation}
q\left( C \right)= \frac{l_{C}}{m} - \left( \frac{d_C}{2m} \right),
\end{equation}
where $C$ denotes a community, $l_C$ denotes the number of the edges connected to the nodes of the community $C$ and $d_C$ represents the sum of the degrees of all of the nodes of $C$. Increase of the value of this function indicates a strong community structure for the nodes of $C$. Now, we can simply replace the $q\left( C \right)$ in (\ref{EQ2}) to re-write the (\ref{EQ14}) as follows:
\begin{singlespace}
\begin{equation}
Q(P)= \sum_{C \in P}{ \frac{l_{C}}{m} - \left( \frac{d_C}{2m} \right)}.
\label{EQ4}
\end{equation}
\end{singlespace}
Using this characteristic of \emph{modularity}, we reduce some of the unnecessary computations by keeping a complementary list for each individual. This list would have a length equal to the number of communities and will contain the \emph{modularity} of each community inside of its elements. Therefore in the case of occurrence of a mutation or crossover, we would be able to calculate the \emph{modularity} of the modified communities only.

After obtaining the \emph{modularity} value for each individual, we use roulette wheel selection method. Using this method, the possibility of selecting an individual is equal to the ratio of its fitness value to the sum of the fitness values of all individuals. This method increases the chance of selecting better individuals based on their fitness scores. Assuming $p_{x}$ as the probability of selecting individual $x$, we can calculate it by:

\begin{equation}
p_{x}=\frac{Q \left( x \right)}{\sum_{y \in \Omega }{Q \left( y \right)}},
\end{equation}
where $\Omega$ indicates the population space. After calculating the probability of each individual, the cumulative probability is computed as follows:
\begin{singlespace}
\begin{equation}
\hat{p_{x}}=\sum_{i=1}^{x} p_{i}.
\end{equation}
\end{singlespace}
Now, if we have a random value between $0$ and $1$ such as $r$, we would select the $k$th individual if and only if $\hat{p}_{k-1}\leq r \leq \hat{p}_k$. In our method, we keep the population space size constant through generations. In each generation, we select $\frac{\vert\Omega\vert}{2}$ couple of individuals using a roulette wheel selection method. Then, we implement a crossover function (section \ref{CROSS}) to produce new individuals. After implementing mutation (section \ref{MUT}) function over a specific ratio of these individuals, we add them to the population space. Finally, we sort the resulted $\vert\Omega\vert+ \frac{\vert\Omega\vert}{2}$ individual based on their fitness values and select the $\vert\Omega\vert$ fittest to transfer to the next generation.

\subsection{Crossover}
\label{CROSS}
\begin{figure*}[t!]
\centerline{
		\includegraphics[width=6in,height=4.5in]{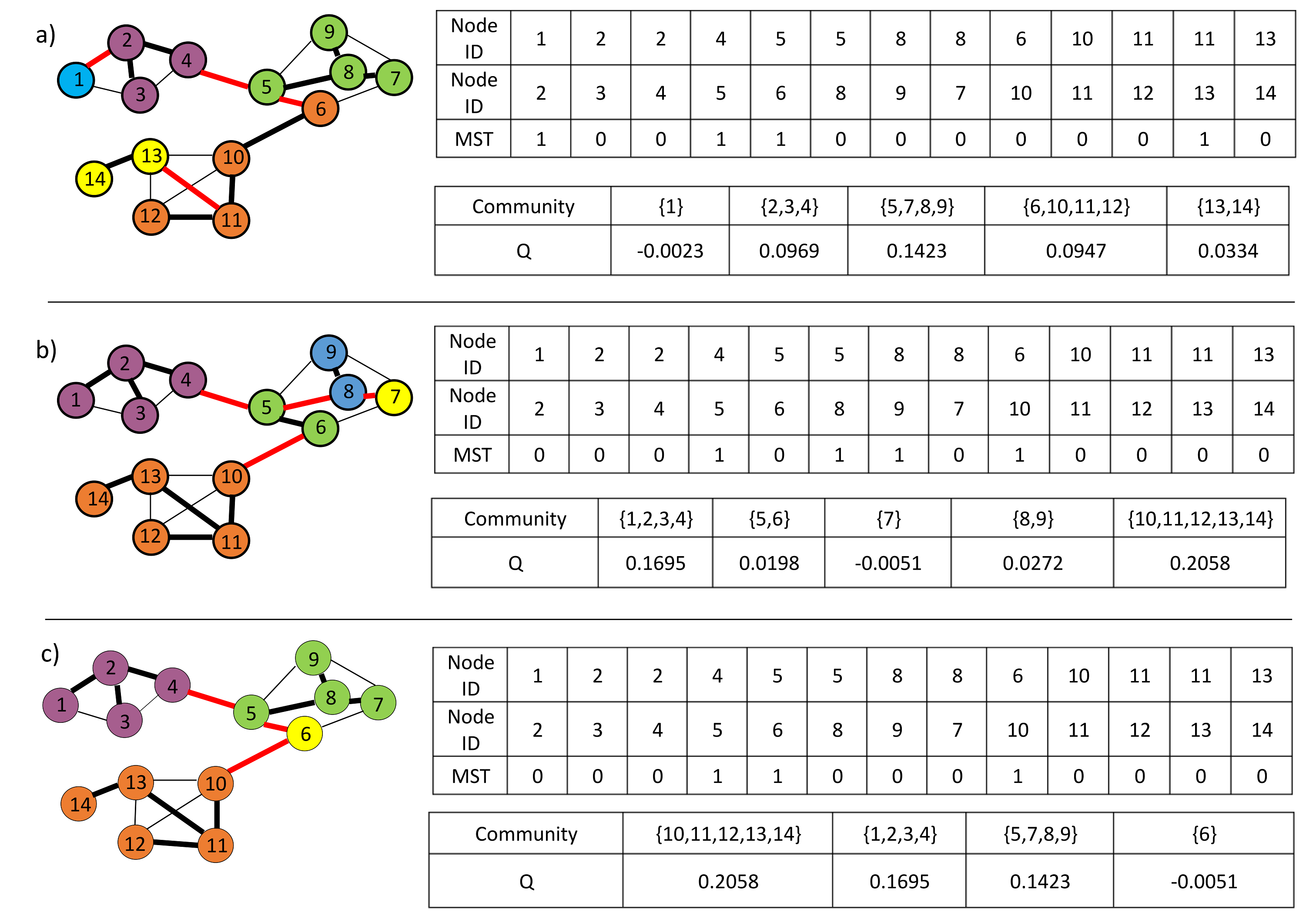}}
	\caption{The crossover procedure to create a new individual; a) The graph, individual and the community vector of the first parent; b) The graph, individual and the community vector of the second parent; c) New individual: to create the new individual first, we sort the communities of both parents in descending order of their modularities, then respectively, we select the communities from best to worst. The $C_{2,5}$, $C_{2,1}$, and $C_{1,3}$ communities get added first. Then $C_{1,4}$ gets selected, but because $\#10$, $\#11$, and $\#12$ nodes are already assigned to other communities, a new singleton community gets formed for the node $\#6$.}
	\label{FIG:3}
\end{figure*}

The crossover function is carried out to merge qualified individuals and create new ones. In \cite{umbarkar2015crossover}, some of the common crossover functions have been reviewed. As we have described in section \ref{SELECT}, we have defined a complementary list to keep the modularity value of each community in hand. Here we use a similar method to \cite{ZALIK201838} as our crossover function. In this method, first, we sort the communities of both parents based on their \emph{modularity} score. Then, we move the corresponding genes of each of these communities to the child individual, respectively.
If for a specific community, part of its genes have already been assigned to the child’s genes by another community, then it gets fragmented. This method iterates through all of the parents’ communities until all of the genes of child individual gets determined.

Figure-\ref{FIG:3} describes the crossover function step by step. It should be noted that by transmitting each community to the child individual, its \emph{modularity} value also gets transmitted unless the community is fragmented. Thus we won't need to re-compute the fitness for each community. Instead, once the child individual was formed, we compute \emph{modularity} only for the fragmented communities by formula (\ref{EQ4}) and sum the values of the rest of the communities.

\subsection{Mutation}
\label{MUT}

\begin{figure}[t]
\begin{subfigure}[t]{2in}
  \centering
  \includegraphics[width=2in,height=2in]{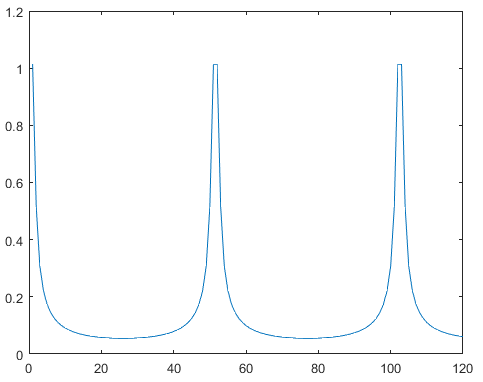}
  \caption{$\alpha=0$}
  \label{fig4:sfig1}
\end{subfigure}%
\begin{subfigure}[t]{2in}
  \centering
  \includegraphics[width=2in,height=2in]{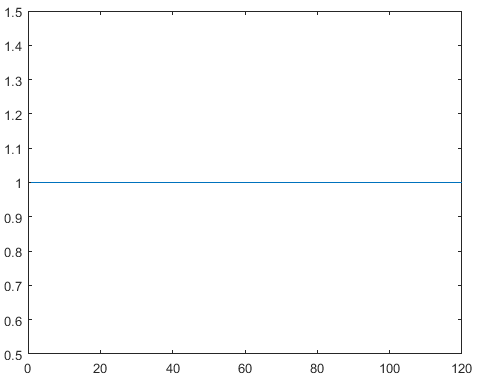}
  \caption{$\alpha=0.5$}
  \label{fig4:sfig2}
\end{subfigure}%
\begin{subfigure}[t]{2in}
  \centering
  \includegraphics[width=2in,height=2in]{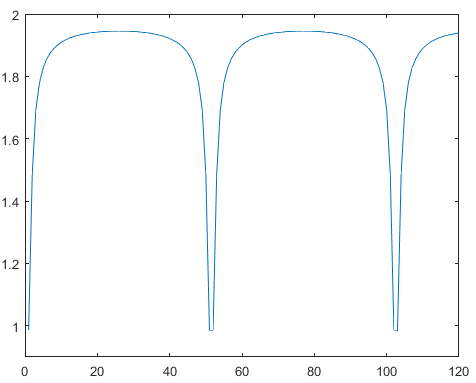}
  \caption{$\alpha=1$}
  \label{fig4:sfig2}
\end{subfigure}
\caption{The effect of $\alpha$ on the shape of probability distribution: a) $\alpha$ parameter has been set equal to $0$ and consequently mutation probability is directed toward border edges; b) setting $\alpha=0.5$, results on uniform probability distribution; c) shows the probability distribution of the edges when setting $\alpha=1$, mutation probability of inner edges have been increased.}
\label{FIG:4}
\end{figure}

A mutation function is used in GAs to add to the diversity of population space. Using this function, one or more genes of each individual gets modified. Here we describe three different mutation functions, two constant and one new adaptive functions.

One of the famous mutation functions is the uniform mutation function. This function assigns an equal chance for each gene to get elected for mutation. This probability would be equal to $\frac{1}{n-1}$ for genes of \emph{MST-based} representation. To select the gene, we create a random number between $1$ and $n-1$. As we have mentioned previously, one of the advantages of the binary representations is the reduction of the value domain for each gene. While in the \emph{locus-based} and \emph{solution-vector} based methods, we have to make another random or educated selection to choose the value of mutated gene (selecting another neighbor in \emph{locus-based} and selecting another community identifier in \emph{solution-vector-based} method), for a binary representation it’s enough to perform a negation.

Also, we have examined a non-uniform mutation function in our experiments. As its name suggests, a non-uniform function does not assign equal probabilities for each gene. We've formed this distribution based on the weights of the corresponding edges of network's MST. To generate the probabilities, we have divided the weight of each edge to the sum of the weights of all edges of the MST. Again we have negated the value of selected genes on mutation. 

\begin{figure*}[t]
\pgfplotsset{width=6in,height=4in}
\begin{tikzpicture}
\begin{axis}[xmax=10,ymax=1.05,xmin=0,ymin=0,domain=0:1,legend pos=north east,xlabel=$z_{q}$, ylabel=$\alpha$]
\addplot[domain=0:10,samples=500] {abs(sin(deg(x)))};
\node[anchor=west] (source1) at (axis cs:0.3,0.2){\parbox{3cm}{\centering $z_0=0.5236$\\ $\vert\sin{z_0}\vert=0.5$}};
\node (destination1) at (axis cs:0.5236,0.5){\textbullet};
\draw[->,ultra thick](source1)--(destination1);

\node[anchor=west] (source2) at (axis cs:0.8,0.5){\parbox{3cm}{\centering $z_1=0.837$\\ $\vert\sin{z_1}\vert=0.7431$}};
\node (destination2) at (axis cs:0.837,0.7431){\textbullet};
\draw[->,ultra thick](source2)--(destination2);

\node[anchor=west] (source3) at (axis cs:2.45,0.8){\parbox{2.8cm}{\centering $z_5=2.09$\\ $\vert\sin{z_5}\vert=0.866$}};
\node (destination3) at (axis cs:2.09,0.866){\textbullet};
\draw[->,ultra thick](source3)--(destination3);

\node[anchor=west] (source3) at (axis cs:4.1,0.3){\parbox{3cm}{\centering $z_9=3.35$\\ $\vert\sin{z_9}\vert=0.2079$}};
\node (destination3) at (axis cs:3.35,0.2079){\textbullet};
\draw[->,ultra thick](source3)--(destination3);
\legend{$\vert\sin(x)\vert$}
\end{axis}
\end{tikzpicture}
\caption{Function of $\alpha_{q}=\vert\sin{z_{q}}\vert$ oscillates between 0 and 1, causing the mutation probability distrubution to change. $\delta$ defines the length of each step ($\delta = 0.1$).}
	\label{FIG:5}
\end{figure*}
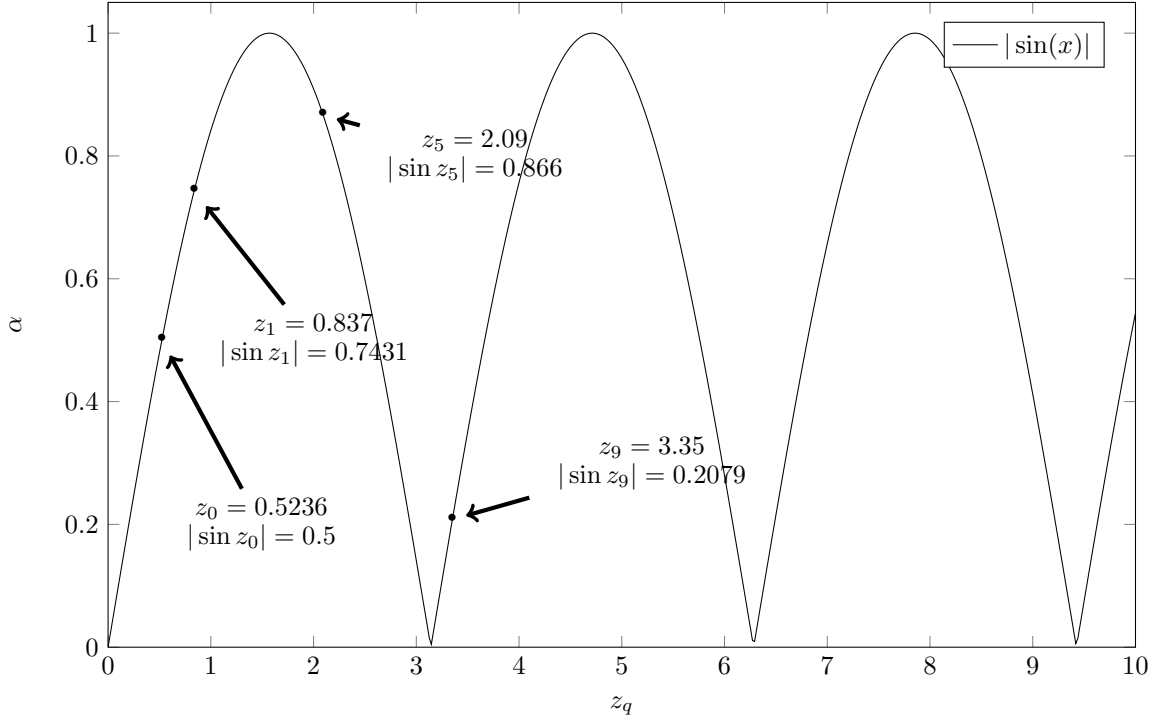

Yet, we also introduce a new adaptive mutation function based on the depths of the corresponding edges of each gene. Naturally, each gene in the \emph{MST-based} representation refers to an edge in the graph’s MST. In some cases, it might be meaningful to keep the internal edges untouched while directing most of the mutations to the border edges or vice versa. Our aim at this function is to examine the effectiveness of this theory. In this method, we assign the probability of each gene concerning its corresponding distance from the \emph{broken edges} of each community. Figure-\ref{FIG:4} can assist us to describe our function. For example, Figure-\ref{FIG:4}.a shows a distribution that puts more weight on the \emph{border edges} (inter-community edges on MST) of the communities. In this figure, those points that have a value of $1$ indicate the \emph{border edges}, and as we move to the depth of each community value gets decreased, then it starts to rise as we move closer to the next \emph{border edge}. Using a distribution similar to Figure-\ref{FIG:4}.a causes to reduce the chance of a mutation on the internal edges. The reverse of that occurs when using a distribution similar to Figure-\ref{FIG:4}.c that internal edges are subjected to the mutations. Figure-\ref{FIG:4}.b displays a situation  that each gene has the same chance to get mutated. To form such a distribution, first, we create a vector such as $\overrightarrow{w}$ with the length of $n-1$ and assign the value of $1$ for the corresponding elements of \emph{broken edges}. Then, for each \emph{broken edge}, we start to navigate the edges connecting to its end nodes in a BFS order until we reach another \emph{broken edge}. For each navigated edge, we change its corresponding $\overrightarrow{w}$ value as follows : 
 \begin{singlespace}
 \begin{equation}
 \overrightarrow{w}_i=\overrightarrow{w}_i + \alpha \cdot 2^{-\frac{1}{d}} + \left(1- \alpha\right)\cdot \left(1- 2^{- \frac{1}{d}} \right).
 \end{equation}
\end{singlespace}
Where the index of the gene is denoted with $i$, $d$ indicates its distance from the \emph{border edge}, and $\alpha$ is a control parameter. Assigning $1$ to $\alpha$ will produce a distribution like Figure-\ref{FIG:4}.c, while $\alpha = 0$ will result in a distribution like Figure-\ref{FIG:4}.a. Whereas assigning $0.5$ to the $\alpha$  would result in a uniform distribution. In our method, $\alpha$ is a self-adaptive parameter. It gets modified based on the \emph{modularity} of the best individual of each generation between $0$ and $1$. If the fitness value of the best individual improves in the last generation, then $\alpha$ remains unchanged, while it changes as follows in the opposite condition:	
\begin{singlespace}
\begin{equation}
\alpha_{q}=
 \begin{cases}
\alpha_{q-1} & \text{if fitness improves} \\
\vert\sin{z_q}\vert & \text{Else}
 \end{cases},
\end{equation}
\end{singlespace}
\begin{equation}
\label{EQ:9}
 z_{q}=\frac{\pi}{6}+q\cdot \delta \cdot \pi,
\end{equation}
where $\delta$ is a user-defined constant to determine the length of each step, and $q$ is the generation index. After performing the described process using all of the \emph{broken edges} of the individuals, we transform the values of the $\overrightarrow{w}$ vector between $0$ and $1$ and use it as the probability distribution function of mutation function. We hypothesize that by assigning the mutation probability of each gene based on its corresponding edge's distance with the \emph{broken edges} of the community, this method can result in mutations that can separate or join the communities in a better way. Furthermore, by adjusting the slope of the mutation probability function based on the improvement of the quality of the best individual of the pool, it can increase the chance of better mutations. Figure-\ref{FIG:5} describes the effect of $\delta$ on determining $\alpha$.

\section{Results and discussion}
\label{RES}
Here we present the results of the proposed method on standard datasets measured based on NMI and modularity, then we compare the results with the results of some other algorithms and discuss the advantages of the proposed method. Both of these measures are widely used measures in community detection literature. Also, we use both real-world and synthetic networks to compare the performance of the proposed method. Finally, we present the results of statistical tests on the results to give a better insight of the algorithms' outcomes.
\subsection{Measures}
\label{MEAS}
NMI and modularity are commonly used measures to qualify the results of community detection algorithms. These measures are defined as follow:

\emph{\textbf{Normalized Mutual Information (NMI)}}: For the datasets that have ground-truth information, it is expected from a community detection algorithm to produce similar outputs to the real-world observations. NMI measures how close the ground-truth and the results are. Taking $\hat{C}=\lbrace \hat{C}_1 , \hat{C}_2 , ... , \hat{C}_k\rbrace$ as the ground-truth data, and $C=\lbrace C_1 , C_2 , ... , C_q\rbrace$ as the detected communities, NMI is calculated as follows:
\begin{equation}
NMI (C, \hat{C})= \frac{2 I(C,\hat{C})}{H(C)+ H(\hat{C})},
\end{equation}
where $I(C,\hat{C})$ and $H(C)$ are the mutual information of $C$ and $\hat{C}$, and the entropy of $C$, respectively, and could be calculated as follows:
\begin{singlespace}
\begin{equation}
I (C, \hat{C})= H(C)+ H(\hat{C})- H(C, \hat{C}),
\end{equation}

\begin{equation}
H(C)= - \sum_{i=1}^{q}{\frac{\vert C_{i} \vert}{n} \log{\frac{\vert C_{i} \vert}{n}}} ,
\end{equation}

\begin{equation}
H(C, \hat{C})= - \sum_{i=1}^{q}{\sum_{j=1}^{k}{\frac{\vert C_{i} \cap C_{j} \vert}{n} \log{\frac{\vert C_{i} \cap C_{j} \vert}{n}}}}.
\end{equation}
\end{singlespace}
\emph{\textbf{Modularity}}: This measure first proposed by Newman and Girvan in \cite{PhysRevE.69.026113} is one of the most popular quality functions applied to measure the quality of a partition. It is defined as follows:
\begin{equation}
Q= \frac{1}{2m} \sum_{i}{\sum_{j}{A_{i,j}-P_{i,j} \delta (C_{i}, C_{j})}},
\label{EQ14}
\end{equation}
where $A$ is the adjacency matrix, $m$ is the number of the edges of the graph, $P$ is a null model matrix that its elements are computed as $P_{i,j}=\frac{\vert \Gamma_{i} \times \Gamma_{j} \vert}{2m}$; $e(i,j) \in E$ , and $\delta(C_i,C_j)$ is a function that returns $1$ whenever $C_i$ and  $C_j$ are the same and $0$ otherwise.

\subsection{Datasets}
Nine real-world and seven synthetic datasets are used to evaluate the performance of the proposed algorithm. Table-\ref{tbl2} summarizes the characteristics of the real-world datasets, and Table-\ref{tbl3} provides the parameters used to generate the synthetic networks. The details of these datasets are described in the following.

\subsubsection{Zachary Karate Club}

This dataset models the interactions between karate club members in an American university. Each node in this network denotes one of the members of the club, and each edge represents the existence of a relationship between two members outside of the club. This network involves $34$ nodes and $78$ edges. The members of the club get divided after a discrepancy between club manager and coach \cite{zachary1977information}.

\subsubsection{Dolphins interaction network}
Bottlenose dolphins were studied in Doubtful Sounds of New Zealand for several years by Lusseau \cite{lusseau2003bottlenose}. $62$ nodes of this network represent the dolphins, and an edge between them indicates that they have been seen together more than expected time.

\subsubsection{American college Football network}
Mostly known as \emph{football network}, this dataset is the result of modeling the matches that took place among the teams of $12$ different conferences in a season of American college football games. On average, each team is more likely to have a match with the teams of its conference ($7$ intra-conference and $4$ inter-conference matches). The ground truth state of this datasets divides the $115$ nodes of the network to $12$ communities in which each community represents a conference \cite{Girvan7821}.

\subsubsection{Political Books}
Amazon.com recommends similar products for its customers while purchasing products. This dataset represents the recommendation network of political books on this website. Books are categorized into three groups based on their content: liberal, neutral, and conservative \cite{krebs2004books}.

\subsubsection{Political Blogs}
Adamic and Glance conducted a study on the internet blogs a few months preceding the 2004 presidential elections of the United States of America and created this dataset. In this graph, each node represents a blog, and an edge takes place between two nodes if they have a hyperlink to each other. Here the aim is to divide the $1490$ node into two communities denoting the political inclination of each blog \cite{10.1145/1134271.1134277}.

\subsubsection{Power Network}

This network shows the power network of western states of the United States of America. In this graph, each node represents a generator, a transformator, or a sub-station, and each edge represents a power supply line between two points. This dataset doesn’t have ground-truth information \cite{watts1998collective}.

\subsubsection{Jazz musicians network}
This dataset represents jazz musicians based on their collaboration. Each of the $2742$ edges of this graph represents the collaboration of jazz musicians in the same band or having a common musician in the band \cite{doi:10.1142/S0219525903001067}.

\subsubsection{Pretty-Good Privacy (PGP)}

This dataset is the network of the users of the Pretty-Good Privacy program, a program that is used for the transmission of encrypted e-mails and files. While each of the 10,680 nodes depicts a user, these nodes are connected by 24,316 edges, each of which depicts a bidirectional signature between two users \cite{PGP2004}.

\subsubsection{Collaboration network of Arxiv High Energy Physics Theory (Ca-HepTh)}

This dataset models the scientific collaboration among the scientists. It covers the papers submitted to the High Energy Physics - Theory category. Each node depicts a scientist, and each edge represents the existence of a co-authored paper. This dataset consists of 9,877 nodes connected by 51,971 edges \cite{caHePh2007}.
\begin{table}%[cols=5,pos=h]%,cols=3,pos=h
\centering
\caption{The summarized information of the real-world datasets used to compare the performance of the proposed method; $N$ indicated the number of nodes of the graph, $m$ denotes the number of edges of the graph, and $\mu$ is the average degree of each node.}\label{tbl2}
\begin{tabular}{c c c c c} 
\hline
Network & N & m & $\mu_d$ & Ground Truth\\
\hline
Karate & $34$ &$78$ & $4.59$ & Y  \\
Dolphins & $62$ &$318$ & $5.13$ & Y  \\
Polbooks & $105$ &$882$ & $8.40$ & Y  \\
Football & $115$ &$1232$ & $10.71$ & Y  \\
Jazz & $198$ &$5484$ & $27.70$ & N  \\
Polblogs & $1490$ &$9545$ & $6.40$ & Y  \\
Power & $4941$ &$13188$ & $2.67$ & N  \\
CA-HepTh & $9877$ &$25998$ & $5.74$ & N  \\
PGP & $10680$ &$24316$ & $4.55$ & N  \\
\hline
\end{tabular}
\end{table}

\begin{table}[t!]%[cols=9,pos=h]
\centering
\caption{Parameters for generating benchmark LFR networks; $N$ indicates the network size, $k_n$ is the average node degree, the maximum node degree is represented with  $k_{max}$, $\mu$ shows the value of mixing parameter, the minimum community size and the maximum community size are denoted with  $C_{min}$ and  $C_{max}$, respectively.}\label{tbl3}
\begin{tabular}{c c c c c c c c c} 
\hline
Network & N & $k_n$ & $k_{max}$ & $\mu$ & $\gamma$ & $\beta$ & $C_{min}$& $C_{max}$\\
\hline
LFR-1 & $50$ &$3$ & $5$ & $0.1$ & $2$& $1$ & $25$ & $25$\\
LFR-2 & $1000$ &$15$ & $20$ & $0.1$ & $2$& $0$ & $500$ & $500$\\
LFR-3 & $1000$ &$20$ & $30$ & $0.1$ & $2$& $1$ & $100$ & $500$\\
LFR-4 & $2500$ &$100$ & $200$ & $0.1$ & $2$& $2$ & $500$ & $1000$\\
LFR-5 & $5000$ &$200$ & $400$ & $0.1$ & $2$& $2$ & $1000$ & $2000$\\
LFR-6 & $10000$ &$100$ & $500$ & $0.1$ & $2$& $2$ & $1000$ & $5000$\\
LFR-$\mu$& $500$ &$20$ & $35$ & $0-0.7$ & $2$& $2$ & $50$ & $100$\\

\hline
\end{tabular}
\end{table}

\subsubsection{Synthetic LFR networks}
\begin{figure*}[!t]
\begin{subfigure}{3in}
  \centering
  \includegraphics[width=3in,height=2.63in]{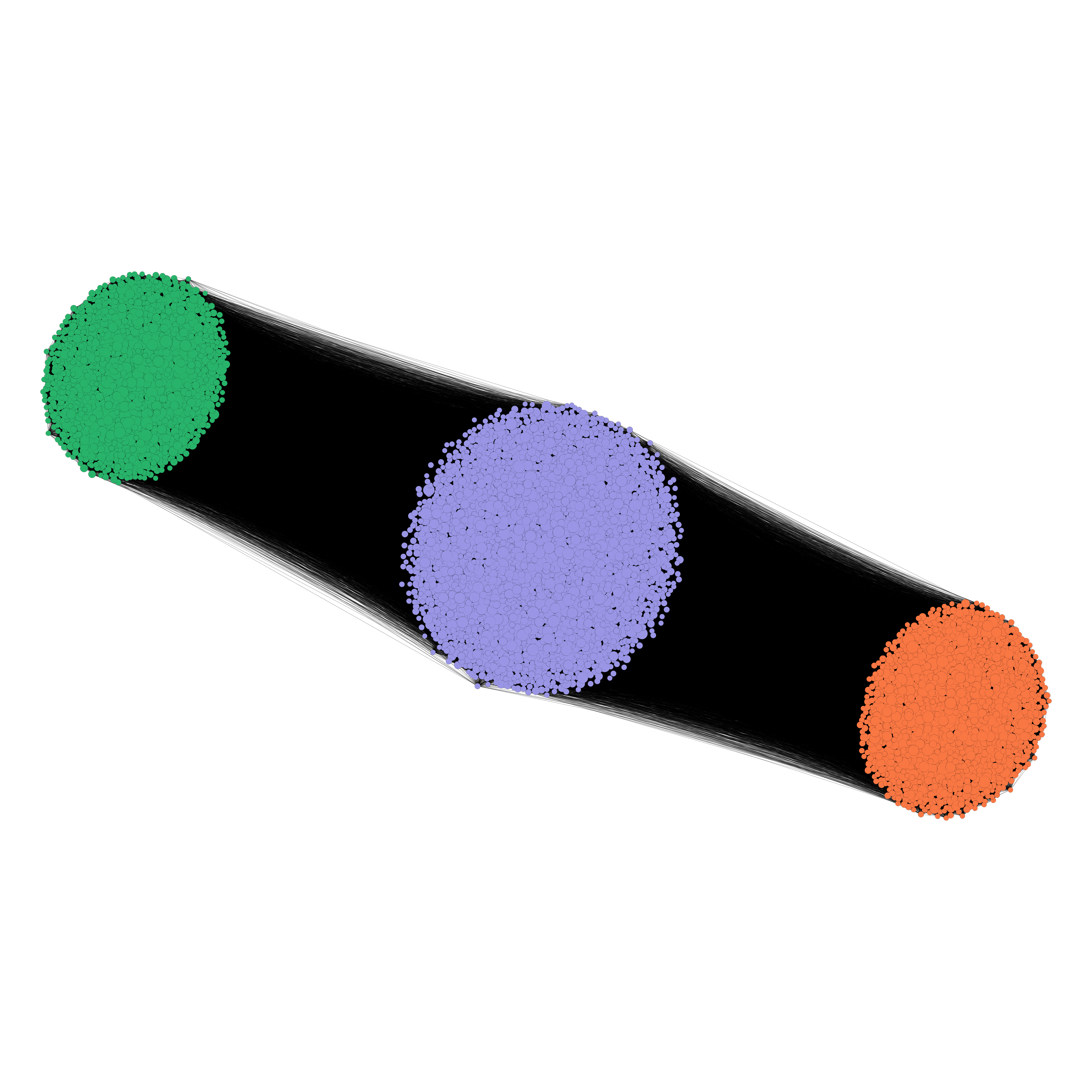}
  \caption{LFR5}
\end{subfigure}%
\begin{subfigure}{3in}
  \centering
  \includegraphics[width=3in,height=2.63in]{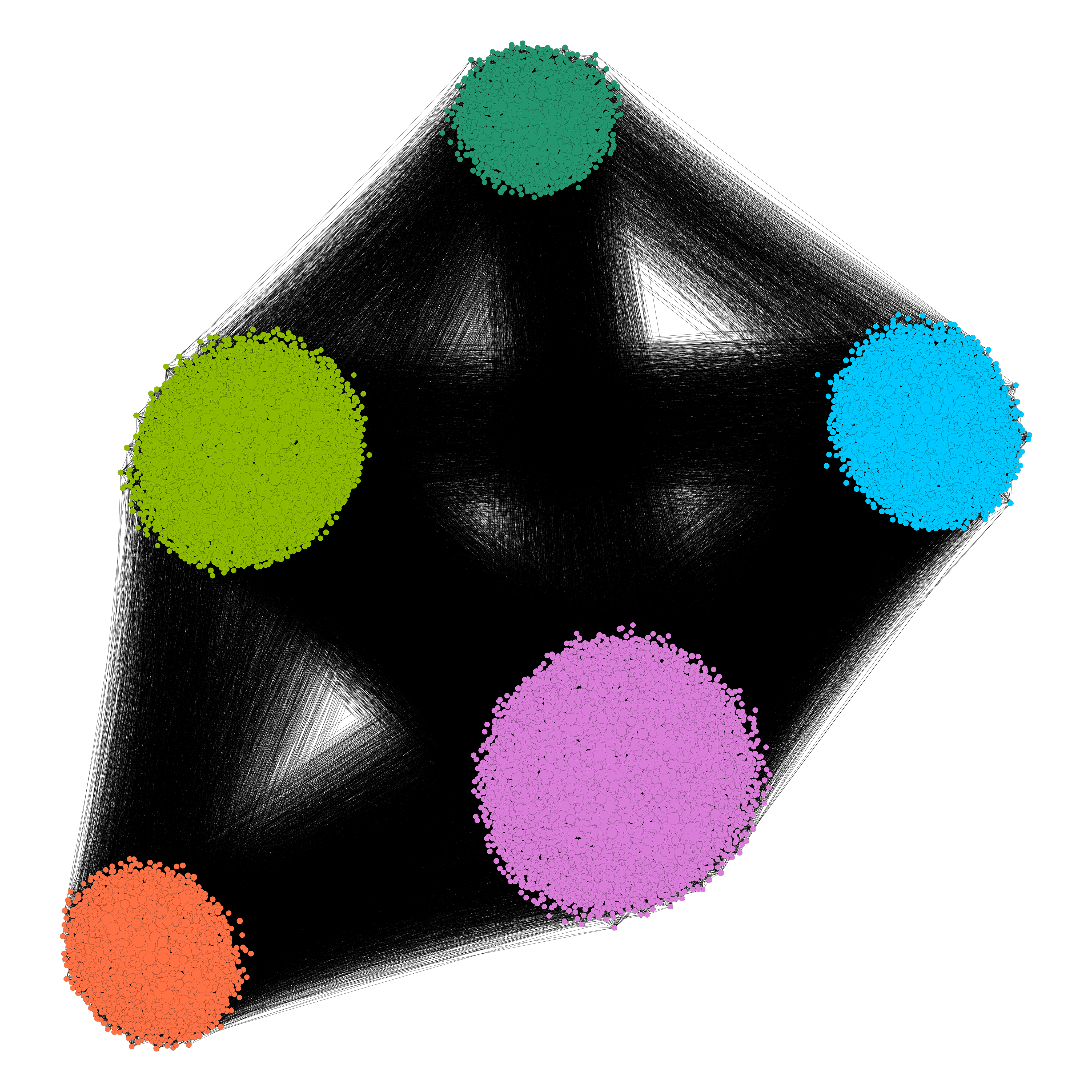}
  \caption{LFR6}
\end{subfigure}%
\caption{Representations of the ground truth partitions of the LFR-5 and LFR-6 networks.}
\label{FIG:6}
\end{figure*}

Lancichinetti, Fortunato, and Radicchi proposed a method to generate the synthetic networks for evaluating the performance of community detection algorithms, which is abbreviated as LFR. In their proposed method, node degree and community size distribution follow a power-law rule and therefore creates networks with similar characteristics of the real networks. Degree and network size exponents are depicted, with $\gamma$ and $\beta$, respectively. Also, they've defined a mixing parameter ($\mu$) that controls the ratio of the edges between communities. We have generated 7 synthetic LFR networks with different sizes for our experiments. The details of the generated networks are given in Table-\ref{tbl3} \cite{lancichinetti2008benchmark}. Figure-\ref{FIG:6} shows two of the synthetic networks generated by the LFR benchmark method for our comparisons.

\subsection{Comparisons}
\label{COMP}
In this section, we evaluate the effect of each of the parameters of the proposed algorithm on the performance of different phases of the proposed algorithm and discuss the advantages and disadvantages of the proposed method compared with other methods. We have conducted extensive performance evaluation experiments and compared the results of the proposed method with some of the recent and classic community detection algorithms. We compared our method with both state-of-the-art and classic algorithms. Also, we picked the algorithms from different categories including, \emph{modularity-optimization-based}, \emph{similarity-optimization-based}, and GA-based methods, to give a better sense of the performances of different algorithms. The compared methods include Louvain \cite{Blondel_2008}, Leiden \cite{Traag2019}, Fastgreedy \cite{PhysRevE.70.066111}, Infomap \cite{Rosvall1118}, LP \cite{PhysRevE.76.036106}, LocalGame \cite{HESAMIPOUR2019122354}, CCGA \cite{said2018cc}, CACD \cite{ying2019parallel}, GA-net \cite{10.1007/978-3-540-87700-4_107}, FluidCom \cite{pares2017fluid}, EdMot \cite{li2019edmot}, WMW \cite{castrillo2017fast}, and WATSET \cite{WATSET}. We have used NMI and modularity measures in our comparisons and conducted the experiments on nine real-world and seven synthetic networks. First, we start by explaining the effects of different internal approaches and parameters. Then we direct the discussion towards the advantages and disadvantages of our method compared to the other algorithms.

\begin{figure*}
\hspace{-0.75in}
\begin{subfigure}[t]{2.5in}
  \centering
  \includegraphics[width=2.5in,height=2.5in]{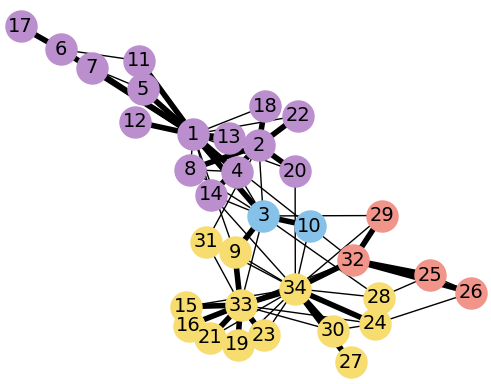}
  \caption{Common Neighbors}
  \label{fig:sfig1}
\end{subfigure}%
\begin{subfigure}[t]{2.5in}
  \centering
  \includegraphics[width=2.5in,height=2.5in]{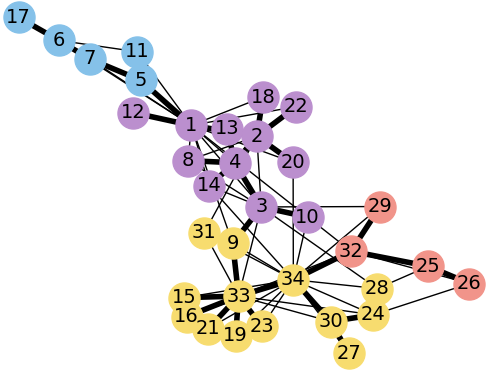}
  \caption{Jaccard}
  \label{fig:sfig2}
\end{subfigure}%
\begin{subfigure}[t]{2.5in}
  \centering
  \includegraphics[width=2.5in,height=2.5in]{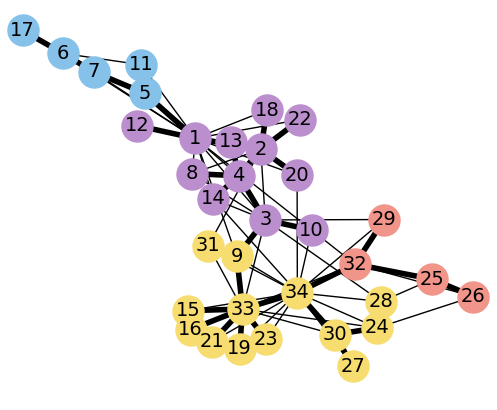}
  \caption{Cosine}
  \label{fig:sfig2}
\end{subfigure}
\\

\hspace{-0.75in}
\begin{subfigure}[t]{2.5in}
  \centering
  \includegraphics[width=2.5in,height=2.5in]{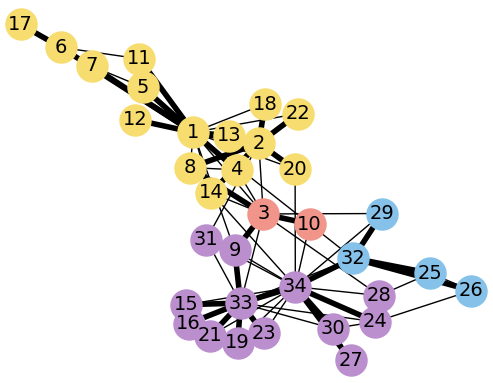}
  \caption{HPI}
  \label{fig:sfig1}
\end{subfigure}%
\begin{subfigure}[t]{2.5in}
  \centering
  \includegraphics[width=2.5in,height=2.5in]{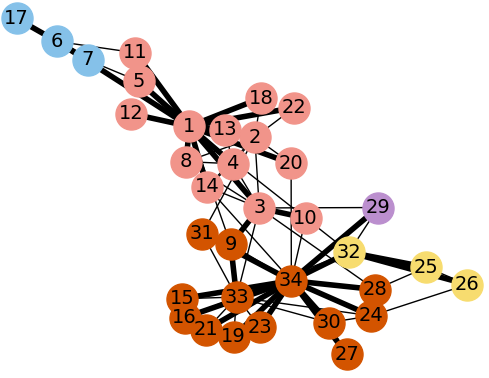}
  \caption{Adamic/Adar}
  \label{fig:sfig2}
\end{subfigure}%
\begin{subfigure}[t]{2.5in}
  \centering
  \includegraphics[width=2.5in,height=2.5in]{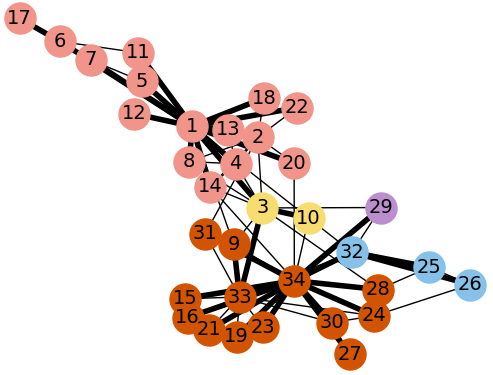}
  \caption{RA}
  \label{fig:sfig2}
\end{subfigure}
\\

\hspace{-0.75in}
\begin{subfigure}[t]{2.5in}
  \centering
  \includegraphics[width=2.5in,height=2.5in]{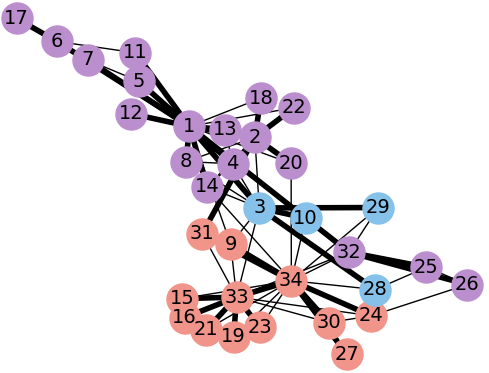}
  \caption{CNDP}
  \label{fig:sfig1}
\end{subfigure}%
\begin{subfigure}[t]{2.5in}
  \centering
  \includegraphics[width=2.5in,height=2.5in]{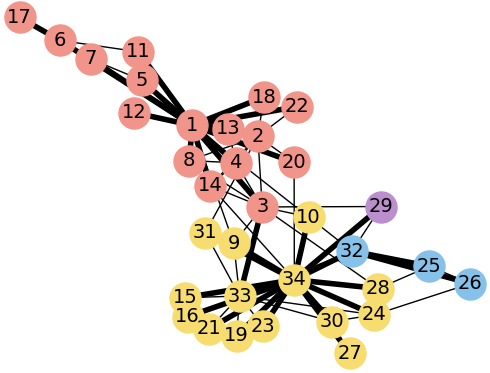}
  \caption{SRW}
  \label{fig:sfig2}
\end{subfigure}%
\begin{subfigure}[t]{2.5in}
  \centering
  \includegraphics[width=2.5in,height=2.5in]{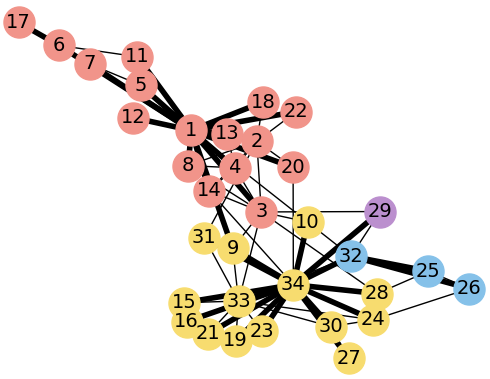}
  \caption{HIN}
  \label{fig:sfig2}
\end{subfigure}

\caption{This figure illustrates the impact of using different similarity measures on the resulted MST tree and the final partition of the algorithm over the Zachary karate club dataset.}
	\label{FIG:7}
\end{figure*}
\begin{figure}[ht]
	\pgfplotsset{width=11cm}
	\centering
 \begin{tikzpicture}
 \begin{axis}[ybar,
 symbolic x coords={CN,Jaccard,Cosine,HPI,AA,RA},
 nodes near coords,
 legend style={at={(0.25,0.95)},
 anchor=north,legend columns=-1},
 nodes near coords align={vertical},]

 \addplot coordinates{(CN,0.7078) (Jaccard,0.6021) (Cosine,0.6021) (HPI,0.7078) (AA,0.5985) (RA,0.6832)};
 \addplot coordinates{(CN,0.3863) (Jaccard,0.4156) (Cosine,0.4156) (HPI,0.3863) (AA,0.3801) (RA,0.3765)};
\legend{NMI,Q}
 \end{axis}
 \end{tikzpicture}
 		\caption{Comparison of the results of the proposed method for different similarity measures in karate network.}
			\label{FIG:8}
\end{figure}
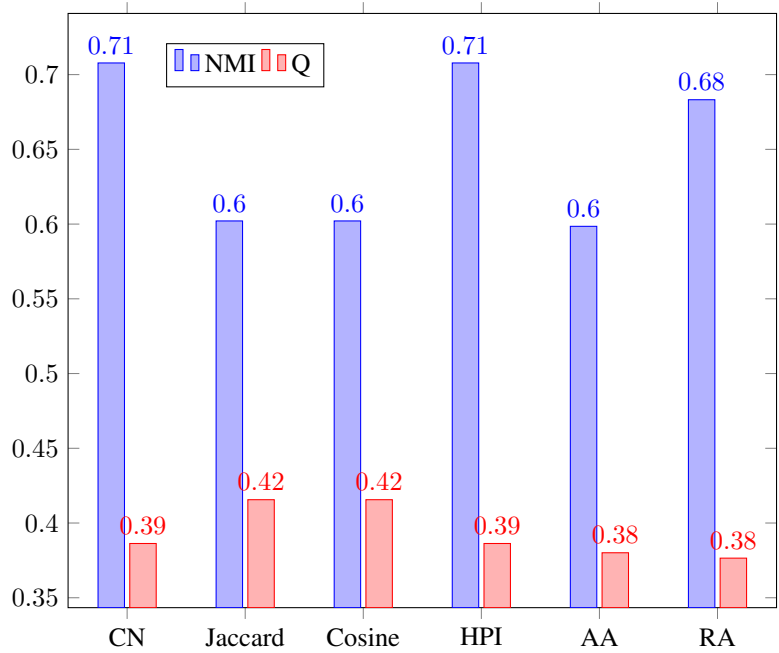

As we have described in section \ref{Rep}, to construct the MST trees needed for the \emph{MST-based} representations, first, we use node similarity measures to assign weight on each edge of the graph. Naturally, based on the applied similarity measure, the resulted MSTs will be different because each measure highlights some characteristics, and therefore some edges might get eliminated when using a specific similarity measure. To determine the best similarity measure, we’ve applied the measures presented in Table-\ref{tbl1} on the karate dataset. Again, we have used different measures from both recent and classic literature and various categories. The resulted MST trees, and the final partitions, are shown in Figure-\ref{FIG:7}. Bolded edges are representing the MST edges. Figure-\ref{FIG:8} compares the results based on the NMI and the modularity. As can be seen from Figure.\ref{FIG:7}, cosine, and Jaccard similarities both perform likewise and better than other similarity measures. It can be noticed in Figure-\ref{FIG:7} that the generated results are directly related to the corresponding MST trees. Specifically considering the edges of node $\#5$, and comparing them with the edges of the same node in different trees, one can understand the impact of each similarity measure. In all of the proceeding experiments, we use \emph{Jaccard} similarity to generate MST trees both because of its simplicity and its higher accuracy.

\noindent
\noindent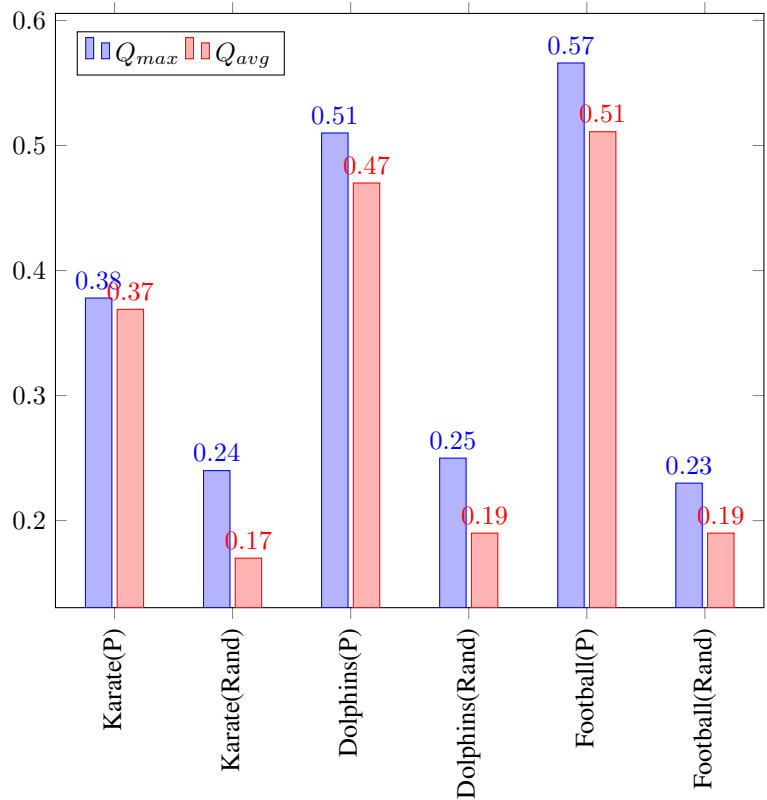
\begin{figure*}[t]
	\centering
	\pgfplotsset{width=11cm}
 \begin{tikzpicture}
 \begin{axis}[ybar,
     x tick label style={rotate=90, anchor=east},
 symbolic x coords={Karate(P), Karate(Rand), Dolphins(P), Dolphins(Rand),Football(P), Football(Rand)},
 %,Jazz(P), Jazz(Rand),Polbooks(P), Polbooks(Rand)
 nodes near coords,
 legend style={legend pos=north west,
legend columns=-1},
 nodes near coords align={vertical},]

\addplot coordinates{(Karate(P),0.378) (Karate(Rand),0.24) (Dolphins(P),0.51) (Dolphins(Rand),0.25) (Football(P),0.566) (Football(Rand),0.23) };
%(Jazz(P),0.356) (Jazz(Rand),0.1) (Polbooks(P),0.5) (Polbooks(Rand),0.21)
 \addplot coordinates{(Karate(P),0.369) (Karate(Rand),0.17) (Dolphins(P),0.47) (Dolphins(Rand),0.19) (Football(P),0.511) (Football(Rand),0.19)};
 %(Jazz(P),0.28) (Jazz(Rand),0.08)(Polbooks(P),0.46) (Polbooks(Rand),0.14)
\legend{$Q_{max}$,$Q_{avg}$}
 \end{axis}
 \end{tikzpicture}
 		\caption{Comparison of the performance of the proposed initial population generation function (P) with random initial population generation function (Rand).}
			\label{FIG:9}
\end{figure*}
\begin{sidewaystable}
\centering
\caption{Comparison of the results of real-world datasets of the proposed method with other methods based on NMI and modularity (Q) measures.}
\label{tbl4}
\begin{tabular}{M{2cm}C{0.6cm}c|C{0.6cm}c|C{0.6cm}c|C{0.6cm}c|C{0.6cm}c|C{0.4cm}C{0cm}|C{0.4cm}C{0cm}|C{0.5cm}C{0cm}|C{0.4cm}C{0cm}|C{0.75cm}C{0.75cm}} 
\hline
 \\[-0.8em]
Network & \multicolumn{2}{c}{Karate} & \multicolumn{2}{c}{Polbooks} & \multicolumn{2}{c}{Polblogs} & \multicolumn{2}{c}{Football} & \multicolumn{2}{c}{Dolphins} &\multicolumn{2}{c}{Power} &  \multicolumn{2}{c}{Jazz} &  \multicolumn{2}{c}{PGP} & \multicolumn{2}{c}{ ca-Hep} &  \multicolumn{2}{c}{Average} \\
 \cline{2-3} \cline{4-5} \cline{6-7} \cline{8-9} \cline{10-11} \cline{12-13} \cline{14-15} \cline{16-17} \cline{18-19} \cline{20-21}
 \\[-0.8em]
 {}&  Q & NMI & Q & NMI & Q & NMI & Q & NMI & Q & NMI & Q & & Q & &  Q & & Q & & $\overline{Q}$ & $\overline{NMI}$ \\

\hline
\textbf{Proposed method} & \textbf{0.42} & $0.60$ & $0.52$ & 0.57 & \textbf{0.43} & $0.45$ &\textbf{0.60} & $0.82$ & \textbf{0.52} & $0.73$ & $0.93$ & & \textbf{0.44} & & 0.86 & &  0.75 & & 0.607 & 0.634\\ 

CACD & \textbf{0.42} & $0.94$ & \textbf{0.53} & \textbf{0.76} & $0.34$ & $0.64$ & $0.58$ & $0.89$ & \textbf{0.52} & \textbf{0.90} & $0.77$ & & $0.43$ & &0.79 & & 0.66 & & 0.56 & \textbf{0.826} \\

CCGA & \textbf{0.42} & $0.69$ & $0.52$ & $0.53$ & $0.37$ & $0.33$ & $0.55$ & $0.70$ & \textbf{0.52} & $0.73$ & $0.90$ & & $0.42$ & & 0.84 & & 0.53 & & 0.563 & 0.596\\

GA-net & \textbf{0.40} & $0.63$ & $0.43$ & $0.40$ & $0.35$ & $0.30$ & $0.60$ & $0.91$ & \textbf{0.42} & $0.58$ & $0.64$ & & $0.37$ & & 0.60 & & 0.53 & & 0.48 & 0.56\\

WATSET & $0.37$ & $1.0$ & $0.49$ & $0.56$ &\textbf{0.43} & $0.48$ & $0.58$ & $0.89$ & $0.48$ & $0.73$ & $0.79$ & & $0.28$ & &0.80 & & 0.69 & & 0.545 & 0.73 \\

FluidCom & $0.35$ & $0.58$ & $0.44$ & $0.48$ &\textbf{0.43} & $0.45$ & $0.58$ & $0.89$ & $0.38$ & $0.46$ & $0.89$ & & $0.42$ & &0.82 & & 0.68 & & 0.554 & 0.572 \\

EdMot & \textbf{0.42} & $0.60$ & \textbf{0.53} & $0.53$ & $0.29$ & $0.36$ & \textbf{0.60} & $0.88$ & \textbf{0.52} & $0.78$ & \textbf{0.94} & & \textbf{0.44} & & 0.87 & &0.77 & & 0.598 & 0.63\\ 

WMW & $0.40$ & $0.54$ & $0.47$ & $0.43$ & $0.40$ & $0.28$ & $0.39$ & $0.69$ & $0.49$ & $0.69$ & $0.61$ & & $0.32$ & &0.67 & & 0.57 & &0.48 & 0.526\\ 
LocalGame & $0.37$ & \textbf{1.0} & $0.51$ & $0.55$ & $0.40$ & $0.68$ & $0.58$ & $0.89$ & $0.51$ & $0.73$ & $0.90$ & & $0.28$ & & 0.79 & & 0.63 & & 0.55 & 0.77\\ 
Leiden & \textbf{0.42} & $0.69$ & $0.52$ & $0.52$ & \textbf{0.43} & $0.63$ & \textbf{0.60} & $0.87$ & \textbf{0.53} & $0.77$ & $0.94$ & & \textbf{0.44} & & \textbf{0.88} & &0.77 & & 0.614 & 0.696\\ 
Louvain & \textbf{0.42} & $0.59$ & $0.52$ & $0.51$ & \textbf{0.43} & $0.63$ & \textbf{0.60} & $0.88$ & \textbf{0.52} & $0.48$ & $0.93$ & & \textbf{0.44} & & \textbf{0.88} & &\textbf{0.82} & & \textbf{0.617} & 0.618\\ 
Fastgreedy & $0.38$ & $0.69$ & $0.50$ & $0.53$ & \textbf{0.43} & $0.65$ & $0.55$ & $0.70$ & $0.50$ & $0.61$ & $0.93$ & & \textbf{0.44} & & 0.85 & & 0.78 & &0.595 & 0.636\\ 
Infomap & $0.40$ & $0.70$ & $0.52$ & $0.49$ & $0.42$ & $0.48$ & \textbf{0.60} & \textbf{0.92} & \textbf{0.52}& $0.50$ & $0.82$ & & $0.28$ & & 0.80 & & 0.73 & &0.565 & 0.618\\ 
LP & $0.40$ & $0.70$ & $0.50$ & 0.57 & \textbf{0.43} & \textbf{0.69} & \textbf{0.60} &\textbf{0.92} & $0.50$ & $0.69$ & $0.81$ & &$0.28$ & & 0.81 & & 0.74 & &0.563 & 0.714\\
%Ground Truth & $0.37$ & $1.0$ & $0.41$ & $1.0$ & $0.41$ & $1.0$ & $0.55$ & $1.0$ & $0.38$ & $1.0$ & $-$ & $-$ & $0.49$ & $1.0$ & $0.43$ & $1.0$ & $0.82$ & $1.0$ & $0.70$ & $1.0$ \\ 
\hline
\end{tabular}
\end{sidewaystable}

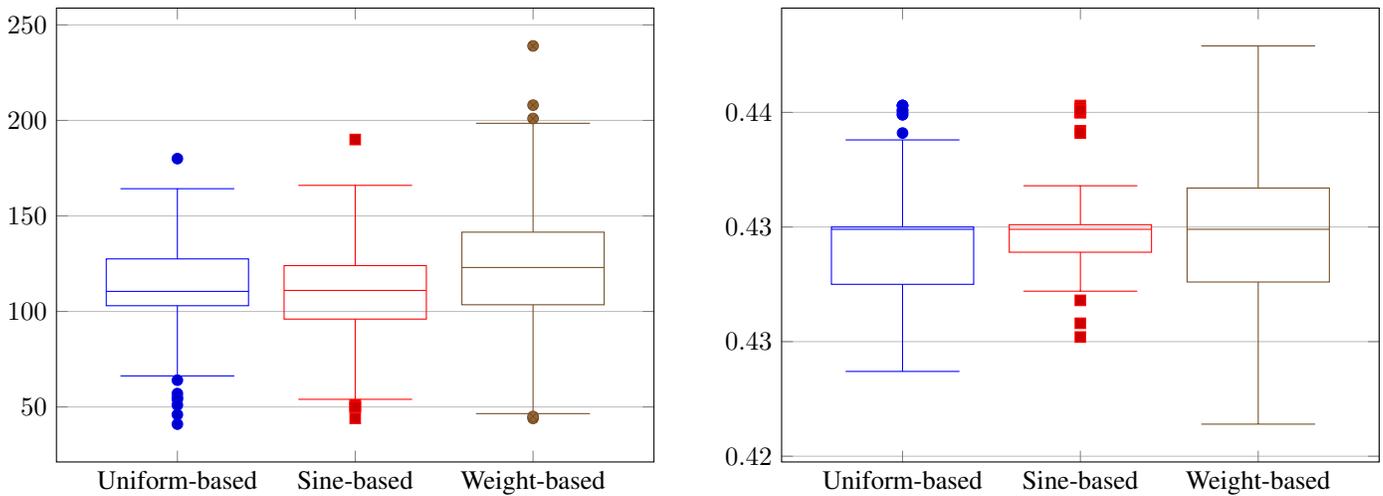
\begin{figure*}[t!]
\hspace{-0.7in}
\begin{subfigure}[t]{3.75in}
\centering
\begin{tikzpicture}[left]
\begin{axis}[
boxplot/draw direction=y,
enlarge y limits,
width=3.75in,
height=3in,
ymajorgrids,
xtick={1,2,3},
xticklabels={Uniform-based, Sine-based, Weight-based},
]
\addplot+ [
boxplot prepared={
lower whisker=66.25, lower quartile=103,
median=110.5,
upper quartile=127.5, upper whisker=164.25,
},
] table [row sep=\\,y index=0] { 64\\ 55\\ 41\\ 57\\ 54\\ 180\\ 51\\ 46\\ }; 
\addplot+ [
boxplot prepared={
lower whisker=54, lower quartile=96,
median=111,
upper quartile=124, upper whisker=166,
},
] table [row sep=\\,y index=0] {51\\ 50\\ 44\\ 50 170\\ 190\\ }; coordinates {};
\addplot+[
boxplot prepared={
lower whisker=46.5, lower quartile=103.5,
median=123,
upper quartile=141.5, upper whisker=198.5,
},
]table [row sep=\\,y index=0] {208\\ 239\\ 201\\ 44\\ 45\\}; coordinates {};
\end{axis}
\end{tikzpicture}
\subcaption{Comparison of mutation functions based on the maximum generation count}
\end{subfigure}
\begin{subfigure}[t]{3.75in}
\centering
\begin{tikzpicture}[right]
\begin{axis}[
boxplot/draw direction=y,
enlarge y limits,
width=3.75in,
height=3in,
ymajorgrids,
xtick={1,2,3},
xticklabels={Uniform-based, Sine-based, Weight-based},
]
\addplot+ [
boxplot prepared={
lower whisker=0.4237, lower quartile=0.4275,
median=0.4299,
upper quartile=0.4300, upper whisker=0.4338,
},
] table [row sep=\\,y index=0] {0.4351\\    0.4353\\    0.4349\\    0.4350\\    0.4349\\    0.4353\\    0.4353\\    0.4341\\    0.4351\\};
\addplot+ [
boxplot prepared={
lower whisker=0.4272, lower quartile=0.4289,
median=0.4299,
upper quartile=0.4301, upper whisker=0.4318,
},
] table [row sep=\\,y index=0] {0.4350\\    0.4351\\    0.4351\\    0.4350\\    0.4342\\    0.4351\\    0.4353\\    0.4351\\    0.4351\\    0.4351\\    0.4351\\    0.4351\\ 0.4351\\    0.4351\\    0.4341\\    0.4350\\    0.4352\\    0.4351\\    0.4351\\    0.4350\\    0.4351\\ 0.4252\\    0.4258\\    0.4268\\}; coordinates {};
\addplot+[
boxplot prepared={
lower whisker=0.4214, lower quartile=0.4276,
median=0.4299,
upper quartile=0.4317, upper whisker=0.4379,
},
]coordinates {};
\end{axis}
\end{tikzpicture}
\subcaption{Comparison of mutation functions based on the modularity score}
\end{subfigure}
\caption{Comparison of variations of the generation count and modularity for 3 different mutation functions. 100 Experiments were performed over the Jazz dataset for each function, while the stopping criteria was, exceeding the modularity of $0.435$, or seeing no improvement on the maximum modularity of each generation on $50$ respective generations; a) Box plots of the number of generations. B) Box plots of the maximum modularity of each experiment.}
	\label{FIG:10}
\end{figure*}

As we explained in Section-\ref{InitPop} we adopt a a very simple initial population generation method by splitting the MST tree into commuities of $\sqrt{n}$ node. Yet this method can result in very effective initial populations which involve near-optimum individuals. To evaluate the abilities of the initial population generation function, we have compared the quality of the populations of a completely randomly generated population genration function with the proposed method. Figure-\ref{FIG:9} compares the modularity of the best individual and the average modularity of the initial population for each approach. It could be seen that the proposed method performs considerably better than the random method. On the other hand in Figure-\ref{FIG:11} generation $0$ depicts the quality of the best individuals of four different methods; 1) the proposed algorithm, 2) CCGA, 3) CACD, and 4) GA-Net. While GA-net's initial population is a random initial population, CACD and CCGA use their own initial population generation function. As it can be seen in Figure-\ref{FIG:11} our initial population generation function produces near-optimum individuals. The method of splliting graph or data to the clusters of $\sqrt{n}$ nodes/points had been implemented already in various clustering and community detection algorithms but as far as we know no other community detection method had implemented this method to create initial population for GA.

The next influencing factor is the mutation function. To examine the effectiveness of the proposed mutation function, we’ve conducted several experiments on the jazz dataset and compared the results based on the convergence time and the modularity score. The obtained results are shown in Figure-\ref{FIG:10}.a. To generate the outcomes, we ran $100$ experiments, while setting the stopping criteria as excessing the modularity of $0.435$ or observing no improvement in the modularity of the best individual for $50$ continuous generations. At first glance, we can see that uncertainty of the number of generations to reach a solution is the highest at the \emph{weight-based} mutation function. This function has a higher mean and longer whiskers than both of the other functions. Also, this method has more upper outliers. Hence we can conclude that this method performs weaker than the others. Considering its higher mean value when compared with the other methods, the main reason for its poor performance comes from the fact that because of its non-uniform probability distribution, it has more tendency to get stuck in some local optima, which might result in higher convergence time. On the other hand, the mean values ($110.5$ and $111$) and the upper outliers of both \emph{sine-based} and \emph{uniform-based} functions are similar. But the box of the \emph{uniform-based} function is more compact than the \emph{sine-based} function's box, and the \emph{sine-based} function's box is slightly inclined lower. Therefore, we can say that the deviations in uniform function are less than the \emph{sine-based} function, and its results are closer to the mean. Yet, it can be observed that the upper whisker of the \emph{sine-based} method is shorter, and its lower whisker is longer than the \emph{uniform-based} method’s whisker, which shows that the \emph{sine-based} function can converge faster than the \emph{uniform-based} method. Therefore, we conclude from the Figure-\ref{FIG:10} that the \emph{sine-based} method performs almost 25\% faster than the \emph{uniform-based} method.

When comparing the variations of the modularity scores of mutation functions, we observe that all of them have a similar mean, but again the \emph{weight-based} method has higher variations compared to the others. Again, this might have resulted out of the non-uniform probability distribution which results in getting stuck in local optima. Therefore, this method performs poorly over both convergence time and the modularity score. Yet, comparing the \emph{sine-based} and \emph{uniform-based} functions shows that both of them have similar upper whiskers, while the lower whisker of the \emph{sine-based} distribution is considerably shorter than the uniform distribution. This means that the likelihood of observing lower modularity is almost 25\% lower when we use a \emph{sine-based} method instead of the \emph{uniform-based}. 
This proves that our hypothesis in Section-\ref{MUT} was right, and the \emph{sine-based} mutation function is able to reduce the convergence time by changing the mutation probability distribution of the genes with a smoothly changing adaptive function. Its higher outcomes come from the fact that \emph{sine-based} method directs the mutation probabilities to the border and the central edges of each community, therefore resulting in joining the smaller communities and breaking the larger ones. Box plots of the modularity variations is shown in the Figure-\ref{FIG:10}.b. 

Consequently, we can conclude that the \emph{sine-based} method can produce better and faster outcomes. It should be considered that this advantage does not come without a cost. For the \emph{sine-based} method, we have to perform a set of calculations, such as assigning a depth value for each edge. Yet, considering that the number of edges in an MST tree is at the order of $n$, then it is possible to compute these values in linear time. Therefore using \emph{sine-based} function can be a cost-effective choice for smaller datasets.

\begin{sidewaystable}

\caption{Comparison of the results of synthetic networks of the proposed method with other methods based on NMI and modularity (Q) measures.}\label{tbl5}

\begin{tabular}{M{2cm}C{0.5cm}c|C{0.5cm}c|C{0.5cm}c|C{0.5cm}c|C{0.6cm}c|C{0.5cm}c|C{0.5cm}c} 
\hline
\\[-0.7em]
Network & \multicolumn{2}{c}{LFR-1} & \multicolumn{2}{c}{LFR-2} & \multicolumn{2}{c}{LFR-3} & \multicolumn{2}{c}{LFR-4} & \multicolumn{2}{c}{LFR-5}& \multicolumn{2}{c}{LFR-6}&\multicolumn{2}{c}{Average} \\ 
 \cline{2-3} \cline{4-5} \cline{6-7} \cline{8-9} \cline{10-11} \cline{12-13} \cline{14-15}
 \\[-0.7em]
 {}& Q & NMI & Q & NMI & Q & NMI & Q & NMI  & Q & NMI & Q & NMI& $\overline{Q}$ & $\overline{NMI}$\\ 
\hline
\textbf{Proposed method} & $0.61$ & $0.54$ & \textbf{0.43} &\textbf{1.0} & \textbf{0.82} & \textbf{1.0} & \textbf{0.70} & \textbf{1.0} & \textbf{0.53} &\textbf{1.0} &\textbf{0.66} &\textbf{1.0} &0.62&0.92 \\ 

CACD & $0.54$ & $0.80$ & \textbf{0.43} & \textbf{1.0} & \textbf{0.82} & \textbf{1.0} & $0.58$ & $0.87$ &0.11&0.57&0.11&0.48 &0.43&0.79 \\ 

CCGA & $0.59$ & $0.56$ & $0.24$ & $0.33$ & $0.64$ & $0.75$ & $0.1$ & $0.37$ &0.1 &0.2 &0.1 &0.1 &0.30&0.39\\ 

GA-net & $0.52$ & $0.43$ & $0.24$ & $0.42$ & $0.61$ & $0.85$ & $0.20$ & $0.30$ &0.1 &0.1 &0.1 &0.1 &0.30&0.37\\ 

WATSET & $0.59$ & $0.65$ & \textbf{0.43} & \textbf{1.0} & \textbf{0.82} & \textbf{1.0} & \textbf{0.70} & \textbf{1.0} &\textbf{0.53} & \textbf{1.0}&\textbf{0.66}&\textbf{1.0} &0.62&0.94 \\ 

FluidCom& $0.54$ & $0.80$ & \textbf{0.43} & \textbf{1.0} & 0.74 & 0.82 & $0.58$ & $0.86$ &0.1&0.44&0.18&0.59&0.42&0.75\\ 

EdMot & \textbf{0.62} & $0.61$ & \textbf{0.43} & \textbf{1.0} & \textbf{0.82} & \textbf{1.0} & \textbf{0.70} & \textbf{1.0} & \textbf{0.53} & \textbf{1.0}& \textbf{0.66} & \textbf{1.0}&0.66&0.93\\

WMW & $0.52$ & $0.43$ & $0.24$ & $0.42$ & $0.74$ & $0.96$ & \textbf{0.70} & \textbf{1.0}&\textbf{0.53} & \textbf{1.0} &\textbf{0.66} &\textbf{1.0}&0.57&0.80 \\ 

LocalGame & $0.49$ & \textbf{1.0} & \textbf{0.43} & \textbf{1.0} & \textbf{0.82} & \textbf{1.0} & \textbf{0.70} & \textbf{1.0} &  \textbf{0.53} & \textbf{1.0} & \textbf{0.66} & \textbf{1.0}&0.60&1.0\\

Leiden & $0.50$ & $0.61$ & \textbf{0.43} & \textbf{1.0} & \textbf{0.82} & \textbf{1.0} & \textbf{0.70} & \textbf{1.0}  & \textbf{0.53} & \textbf{1.0} &\textbf{0.66} &\textbf{1.0}&0.61&0.93 \\ 

Louvain & $0.47$ & $0.39$ & \textbf{0.43} & \textbf{1.0} & \textbf{0.82} & \textbf{1.0} & \textbf{0.70} & \textbf{1.0}  & \textbf{0.53} & \textbf{1.0} &\textbf{0.66} &\textbf{1.0}&0.60&0.90\\ 
Fastgreedy & \textbf{0.62} & $0.66$ & \textbf{0.43} & \textbf{1.0} & $0.81$ & $0.99$ & $0.69$ & $0.99$ & \textbf{0.53} & \textbf{1.0} &\textbf{0.66} &\textbf{1.0}&0.62&0.94\\ 
Infomap & $0.60$ & $0.54$ & \textbf{0.43} & \textbf{1.0} & \textbf{0.82} & \textbf{1.0} & \textbf{0.70} & \textbf{1.0}& \textbf{0.53} & \textbf{1.0} &\textbf{0.66} &\textbf{1.0}&0.62&0.92\\ 
LP & $0.56$ & $0.80$ & \textbf{0.43} & \textbf{1.0} & \textbf{0.82} & \textbf{1.0} & \textbf{0.70} & \textbf{1.0} & \textbf{0.53} & \textbf{1.0} &\textbf{0.66} &\textbf{1.0} &0.62&0.97\\  
%Ground Truth & $0.37$ & $1.0$ & $0.41$ & $1.0$ & $0.41$ & $1.0$ & $0.55$ & $1.0$ & $0.38$ & $1.0$ & $-$ & $-$ & $0.49$ & $1.0$ & $0.43$ & $1.0$ & $0.82$ & $1.0$ & $0.70$ & $1.0$ \\ 
\hline
\end{tabular}
\end{sidewaystable}

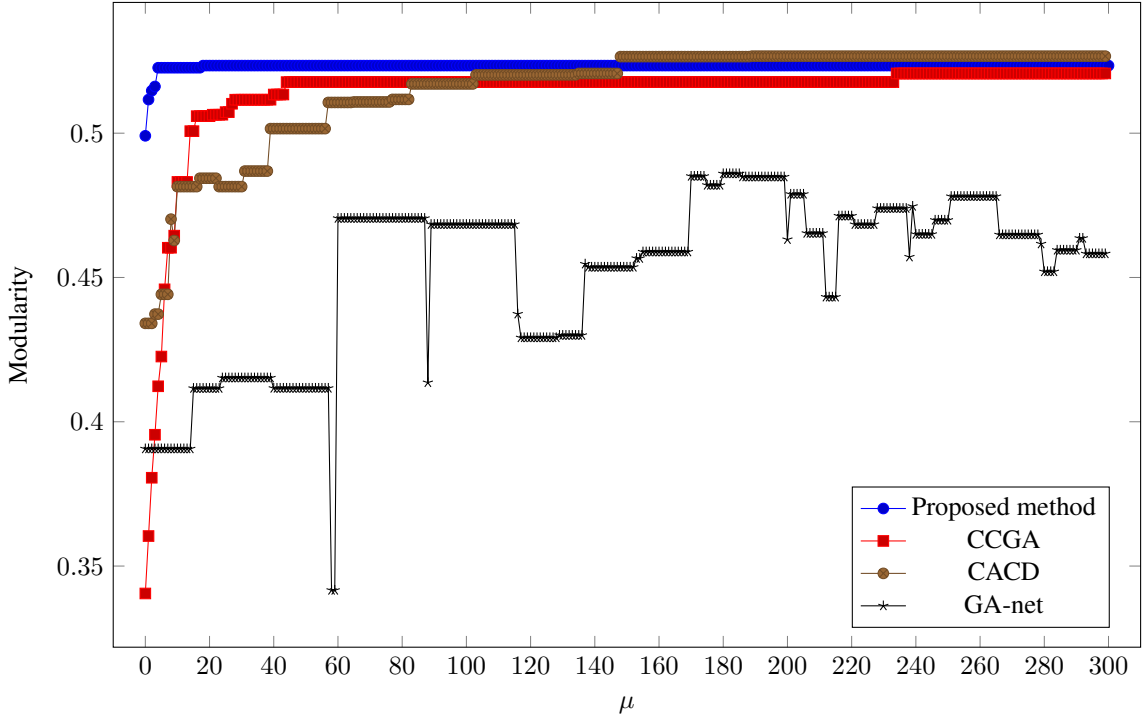
\begin{figure*}[t!]
\begin{tikzpicture}
\begin{axis}[
width=6in,
height=4in,
xmin=-10,
xmax=310,
legend style={legend pos=south east},
xlabel=$\mu$, ylabel=Modularity]
%GenMST
\addplot coordinates {
 (0,0.49909) (1,0.51165) (2,0.51472) (3,0.51618) (4,0.52268) (5,0.52268) (6,0.52268) (7,0.52268) (8,0.52268) (9,0.52268) (10,0.52268) (11,0.52268) (12,0.52268) (13,0.52268) (14,0.52268) (15,0.52268) (16,0.52268) (17,0.52268) (18,0.52342) (19,0.52342) (20,0.52342) (21,0.52342) (22,0.52342) (23,0.52342) (24,0.52342) (25,0.52342) (26,0.52342) (27,0.52342) (28,0.52342) (29,0.52342) (30,0.52342) (31,0.52342) (32,0.52342) (33,0.52342) (34,0.52342) (35,0.52342) (36,0.52342) (37,0.52342) (38,0.52342) (39,0.52342) (40,0.52342) (41,0.52342) (42,0.52342) (43,0.52342) (44,0.52342) (45,0.52342) (46,0.52342) (47,0.52342) (48,0.52342) (49,0.52342) (50,0.52342) (51,0.52342) (52,0.52342) (53,0.52342) (54,0.52342) (55,0.52342) (56,0.52342) (57,0.52342) (58,0.52342) (59,0.52342) (60,0.52342) (61,0.52342) (62,0.52342) (63,0.52342) (64,0.52342) (65,0.52342) (66,0.52342) (67,0.52342) (68,0.52342) (69,0.52342) (70,0.52342) (71,0.52342) (72,0.52342) (73,0.52342) (74,0.52342) (75,0.52342) (76,0.52342) (77,0.52342) (78,0.52342) (79,0.52342) (80,0.52342) (81,0.52342) (82,0.52342) (83,0.52342) (84,0.52342) (85,0.52342) (86,0.52342) (87,0.52342) (88,0.52342) (89,0.52342) (90,0.52342) (91,0.52342) (92,0.52342) (93,0.52342) (94,0.52342) (95,0.52342) (96,0.52342) (97,0.52342) (98,0.52342) (99,0.52342) (100,0.52342) (101,0.52342) (102,0.52342) (103,0.52342) (104,0.52342) (105,0.52342) (106,0.52342) (107,0.52342) (108,0.52342) (109,0.52342) (110,0.52342) (111,0.52342) (112,0.52342) (113,0.52342) (114,0.52342) (115,0.52342) (116,0.52342) (117,0.52342) (118,0.52342) (119,0.52342) (120,0.52342) (121,0.52342) (122,0.52342) (123,0.52342) (124,0.52342) (125,0.52342) (126,0.52342) (127,0.52342) (128,0.52342) (129,0.52342) (130,0.52342) (131,0.52342) (132,0.52342) (133,0.52342) (134,0.52342) (135,0.52342) (136,0.52342) (137,0.52342) (138,0.52342) (139,0.52342) (140,0.52342) (141,0.52342) (142,0.52342) (143,0.52342) (144,0.52342) (145,0.52342) (146,0.52342) (147,0.52342) (148,0.52342) (149,0.52342) (150,0.52342) (151,0.52342) (152,0.52342) (153,0.52342) (154,0.52342) (155,0.52342) (156,0.52342) (157,0.52342) (158,0.52342) (159,0.52342) (160,0.52342) (161,0.52342) (162,0.52342) (163,0.52342) (164,0.52342) (165,0.52342) (166,0.52342) (167,0.52342) (168,0.52342) (169,0.52342) (170,0.52342) (171,0.52342) (172,0.52342) (173,0.52342) (174,0.52342) (175,0.52342) (176,0.52342) (177,0.52342) (178,0.52342) (179,0.52342) (180,0.52342) (181,0.52342) (182,0.52342) (183,0.52342) (184,0.52342) (185,0.52342) (186,0.52342) (187,0.52342) (188,0.52342) (189,0.52342) (190,0.52342) (191,0.52342) (192,0.52342) (193,0.52342) (194,0.52342) (195,0.52342) (196,0.52342) (197,0.52342) (198,0.52342) (199,0.52342) (200,0.52342) (201,0.52342) (202,0.52342) (203,0.52342) (204,0.52342) (205,0.52342) (206,0.52342) (207,0.52342) (208,0.52342) (209,0.52342) (210,0.52342) (211,0.52342) (212,0.52342) (213,0.52342) (214,0.52342) (215,0.52342) (216,0.52342) (217,0.52342) (218,0.52342) (219,0.52342) (220,0.52342) (221,0.52342) (222,0.52342) (223,0.52342) (224,0.52342) (225,0.52342) (226,0.52342) (227,0.52342) (228,0.52342) (229,0.52342) (230,0.52342) (231,0.52342) (232,0.52342) (233,0.52342) (234,0.52342) (235,0.52342) (236,0.52342) (237,0.52342) (238,0.52342) (239,0.52342) (240,0.52342) (241,0.52342) (242,0.52342) (243,0.52342) (244,0.52342) (245,0.52342) (246,0.52342) (247,0.52342) (248,0.52342) (249,0.52342) (250,0.52342) (251,0.52342) (252,0.52342) (253,0.52342) (254,0.52342) (255,0.52342) (256,0.52342) (257,0.52342) (258,0.52342) (259,0.52342) (260,0.52342) (261,0.52342) (262,0.52342) (263,0.52342) (264,0.52342) (265,0.52342) (266,0.52342) (267,0.52342) (268,0.52342) (269,0.52342) (270,0.52342) (271,0.52342) (272,0.52342) (273,0.52342) (274,0.52342) (275,0.52342) (276,0.52342) (277,0.52342) (278,0.52342) (279,0.52342) (280,0.52342) (281,0.52342) (282,0.52342) (283,0.52342) (284,0.52342) (285,0.52342) (286,0.52342) (287,0.52342) (288,0.52342) (289,0.52342) (290,0.52342) (291,0.52342) (292,0.52342) (293,0.52342) (294,0.52342) (295,0.52342) (296,0.52342) (297,0.52342) (298,0.52342) (299,0.52342) (300,0.52342)};
 %CCGA
 \addplot coordinates{ (0,0.3405) (1,0.3604) (2,0.3806) (3,0.3955) (4,0.4123) (5,0.4226) (6,0.4459) (7,0.4603) (8,0.4603) (9,0.4645) (10,0.4831) (11,0.4831) (12,0.4831) (13,0.4831) (14,0.5007) (15,0.5007) (16,0.5059) (17,0.5059) (18,0.5059) (19,0.5059) (20,0.5059) (21,0.5064) (22,0.5064) (23,0.5064) (24,0.5064) (25,0.5073) (26,0.5073) (27,0.5102) (28,0.5116) (29,0.5116) (30,0.5116) (31,0.5116) (32,0.5116) (33,0.5116) (34,0.5116) (35,0.5116) (36,0.5116) (37,0.5116) (38,0.5116) (39,0.5116) (40,0.5134) (41,0.5134) (42,0.5134) (43,0.5134) (44,0.5177) (45,0.5177) (46,0.5177) (47,0.5177) (48,0.5177) (49,0.5177) (50,0.5177) (51,0.5177) (52,0.5177) (53,0.5177) (54,0.5177) (55,0.5177) (56,0.5177) (57,0.5177) (58,0.5177) (59,0.5177) (60,0.5177) (61,0.5177) (62,0.5177) (63,0.5177) (64,0.5177) (65,0.5177) (66,0.5177) (67,0.5177) (68,0.5177) (69,0.5177) (70,0.5177) (71,0.5177) (72,0.5177) (73,0.5177) (74,0.5177) (75,0.5177) (76,0.5177) (77,0.5177) (78,0.5177) (79,0.5177) (80,0.5177) (81,0.5177) (82,0.5177) (83,0.5177) (84,0.5177) (85,0.5177) (86,0.5177) (87,0.5177) (88,0.5177) (89,0.5177) (90,0.5177) (91,0.5177) (92,0.5177) (93,0.5177) (94,0.5177) (95,0.5177) (96,0.5177) (97,0.5177) (98,0.5177) (99,0.5177) (100,0.5177) (101,0.5177) (102,0.5177) (103,0.5177) (104,0.5177) (105,0.5177) (106,0.5177) (107,0.5177) (108,0.5177) (109,0.5177) (110,0.5177) (111,0.5177) (112,0.5177) (113,0.5177) (114,0.5177) (115,0.5177) (116,0.5177) (117,0.5177) (118,0.5177) (119,0.5177) (120,0.5177) (121,0.5177) (122,0.5177) (123,0.5177) (124,0.5177) (125,0.5177) (126,0.5177) (127,0.5177) (128,0.5177) (129,0.5177) (130,0.5177) (131,0.5177) (132,0.5177) (133,0.5177) (134,0.5177) (135,0.5177) (136,0.5177) (137,0.5177) (138,0.5177) (139,0.5177) (140,0.5177) (141,0.5177) (142,0.5177) (143,0.5177) (144,0.5177) (145,0.5177) (146,0.5177) (147,0.5177) (148,0.5177) (149,0.5177) (150,0.5177) (151,0.5177) (152,0.5177) (153,0.5177) (154,0.5177) (155,0.5177) (156,0.5177) (157,0.5177) (158,0.5177) (159,0.5177) (160,0.5177) (161,0.5177) (162,0.5177) (163,0.5177) (164,0.5177) (165,0.5177) (166,0.5177) (167,0.5177) (168,0.5177) (169,0.5177) (170,0.5177) (171,0.5177) (172,0.5177) (173,0.5177) (174,0.5177) (175,0.5177) (176,0.5177) (177,0.5177) (178,0.5177) (179,0.5177) (180,0.5177) (181,0.5177) (182,0.5177) (183,0.5177) (184,0.5177) (185,0.5177) (186,0.5177) (187,0.5177) (188,0.5177) (189,0.5177) (190,0.5177) (191,0.5177) (192,0.5177) (193,0.5177) (194,0.5177) (195,0.5177) (196,0.5177) (197,0.5177) (198,0.5177) (199,0.5177) (200,0.5177) (201,0.5177) (202,0.5177) (203,0.5177) (204,0.5177) (205,0.5177) (206,0.5177) (207,0.5177) (208,0.5177) (209,0.5177) (210,0.5177) (211,0.5177) (212,0.5177) (213,0.5177) (214,0.5177) (215,0.5177) (216,0.5177) (217,0.5177) (218,0.5177) (219,0.5177) (220,0.5177) (221,0.5177) (222,0.5177) (223,0.5177) (224,0.5177) (225,0.5177) (226,0.5177) (227,0.5177) (228,0.5177) (229,0.5177) (230,0.5177) (231,0.5177) (232,0.5177) (233,0.5177) (234,0.5207) (235,0.5207) (236,0.5207) (237,0.5207) (238,0.5207) (239,0.5207) (240,0.5207) (241,0.5207) (242,0.5207) (243,0.5207) (244,0.5207) (245,0.5207) (246,0.5207) (247,0.5207) (248,0.5207) (249,0.5207) (250,0.5207) (251,0.5207) (252,0.5207) (253,0.5207) (254,0.5207) (255,0.5207) (256,0.5207) (257,0.5207) (258,0.5207) (259,0.5207) (260,0.5207) (261,0.5207) (262,0.5207) (263,0.5207) (264,0.5207) (265,0.5207) (266,0.5207) (267,0.5207) (268,0.5207) (269,0.5207) (270,0.5207) (271,0.5207) (272,0.5207) (273,0.5207) (274,0.5207) (275,0.5207) (276,0.5207) (277,0.5207) (278,0.5207) (279,0.5207) (280,0.5207) (281,0.5207) (282,0.5207) (283,0.5207) (284,0.5207) (285,0.5207) (286,0.5207) (287,0.5207) (288,0.5207) (289,0.5207) (290,0.5207) (291,0.5207) (292,0.5207) (293,0.5207) (294,0.5207) (295,0.5207) (296,0.5207) (297,0.5207) (298,0.5207) (299,0.5207)};
  \addplot coordinates{(0,0.4341) (1,0.4341) (2,0.4341) (3,0.4373) (4,0.4373) (5,0.44416) (6,0.44416) (7,0.44416) (8,0.47021) (9,0.46281) (10,0.48148) (11,0.48148) (12,0.48148) (13,0.48148) (14,0.48148) (15,0.48148) (16,0.48148) (17,0.48439) (18,0.48439) (19,0.48439) (20,0.48439) (21,0.48439) (22,0.48439) (23,0.48148) (24,0.48148) (25,0.48148) (26,0.48148) (27,0.48148) (28,0.48148) (29,0.48148) (30,0.48148) (31,0.48687) (32,0.48687) (33,0.48687) (34,0.48687) (35,0.48687) (36,0.48687) (37,0.48687) (38,0.48687) (39,0.50158) (40,0.50158) (41,0.50158) (42,0.50158) (43,0.50158) (44,0.50158) (45,0.50158) (46,0.50158) (47,0.50158) (48,0.50158) (49,0.50158) (50,0.50158) (51,0.50158) (52,0.50158) (53,0.50158) (54,0.50158) (55,0.50158) (56,0.50158) (57,0.51065) (58,0.51065) (59,0.51065) (60,0.51065) (61,0.51065) (62,0.51065) (63,0.51065) (64,0.51065) (65,0.51082) (66,0.51082) (67,0.51082) (68,0.51082) (69,0.51082) (70,0.51082) (71,0.51082) (72,0.51082) (73,0.51082) (74,0.51082) (75,0.51082) (76,0.51082) (77,0.51171) (78,0.51171) (79,0.51171) (80,0.51171) (81,0.51171) (82,0.51171) (83,0.51707) (84,0.51707) (85,0.51707) (86,0.51707) (87,0.51707) (88,0.51707) (89,0.51707) (90,0.51707) (91,0.51707) (92,0.51707) (93,0.51707) (94,0.51707) (95,0.51707) (96,0.51707) (97,0.51707) (98,0.51707) (99,0.51707) (100,0.51707) (101,0.51707) (102,0.51707) (103,0.52019) (104,0.52019) (105,0.52019) (106,0.52019) (107,0.52019) (108,0.52019) (109,0.52019) (110,0.52019) (111,0.52019) (112,0.52019) (113,0.52019) (114,0.52019) (115,0.52019) (116,0.52019) (117,0.52019) (118,0.52019) (119,0.52019) (120,0.52019) (121,0.52019) (122,0.52019) (123,0.52022) (124,0.52022) (125,0.52022) (126,0.52022) (127,0.52022) (128,0.52022) (129,0.52022) (130,0.52022) (131,0.52022) (132,0.52022) (133,0.52022) (134,0.52022) (135,0.52069) (136,0.52069) (137,0.52069) (138,0.52069) (139,0.52069) (140,0.52069) (141,0.52069) (142,0.52069) (143,0.52069) (144,0.52069) (145,0.52069) (146,0.52069) (147,0.52069) (148,0.52653) (149,0.52653) (150,0.52653) (151,0.52653) (152,0.52653) (153,0.52653) (154,0.52653) (155,0.52653) (156,0.52653) (157,0.52653) (158,0.52653) (159,0.52653) (160,0.52653) (161,0.52653) (162,0.52653) (163,0.52653) (164,0.52653) (165,0.52653) (166,0.52653) (167,0.52653) (168,0.52653) (169,0.52653) (170,0.52653) (171,0.52653) (172,0.52653) (173,0.52653) (174,0.52653) (175,0.52653) (176,0.52653) (177,0.52653) (178,0.52653) (179,0.52653) (180,0.52653) (181,0.52653) (182,0.52653) (183,0.52653) (184,0.52653) (185,0.52653) (186,0.52653) (187,0.52653) (188,0.52653) (189,0.52669) (190,0.52669) (191,0.52669) (192,0.52669) (193,0.52669) (194,0.52669) (195,0.52669) (196,0.52669) (197,0.52669) (198,0.52669) (199,0.52669) (200,0.52669) (201,0.52669) (202,0.52669) (203,0.52669) (204,0.52669) (205,0.52669) (206,0.52669) (207,0.52669) (208,0.52669) (209,0.52669) (210,0.52669) (211,0.52669) (212,0.52669) (213,0.52669) (214,0.52669) (215,0.52669) (216,0.52669) (217,0.52669) (218,0.52669) (219,0.52669) (220,0.52669) (221,0.52669) (222,0.52669) (223,0.52669) (224,0.52669) (225,0.52669) (226,0.52669) (227,0.52669) (228,0.52669) (229,0.52669) (230,0.52669) (231,0.52669) (232,0.52669) (233,0.52669) (234,0.52669) (235,0.52669) (236,0.52669) (237,0.52669) (238,0.52669) (239,0.52669) (240,0.52669) (241,0.52669) (242,0.52669) (243,0.52669) (244,0.52669) (245,0.52669) (246,0.52669) (247,0.52669) (248,0.52669) (249,0.52669) (250,0.52669) (251,0.52669) (252,0.52669) (253,0.52669) (254,0.52669) (255,0.52669) (256,0.52669) (257,0.52669) (258,0.52669) (259,0.52669) (260,0.52669) (261,0.52669) (262,0.52669) (263,0.52669) (264,0.52669) (265,0.52669) (266,0.52669) (267,0.52669) (268,0.52669) (269,0.52669) (270,0.52669) (271,0.52669) (272,0.52669) (273,0.52669) (274,0.52669) (275,0.52669) (276,0.52669) (277,0.52669) (278,0.52669) (279,0.52669) (280,0.52669) (281,0.52669) (282,0.52669) (283,0.52669) (284,0.52669) (285,0.52669) (286,0.52669) (287,0.52669) (288,0.52669) (289,0.52669) (290,0.52669) (291,0.52669) (292,0.52669) (293,0.52669) (294,0.52669) (295,0.52669) (296,0.52669) (297,0.52669) (298,0.52669) (299,0.52669) };
   \addplot coordinates{(0,0.39062) (1,0.39062) (2,0.39062) (3,0.39062) (4,0.39062) (5,0.39062) (6,0.39062) (7,0.39062) (8,0.39062) (9,0.39062) (10,0.39062) (11,0.39062) (12,0.39062) (13,0.39062) (14,0.39062) (15,0.41157) (16,0.41157) (17,0.41157) (18,0.41157) (19,0.41157) (20,0.41157) (21,0.41157) (22,0.41157) (23,0.41157) (24,0.41519) (25,0.41519) (26,0.41519) (27,0.41519) (28,0.41519) (29,0.41519) (30,0.41519) (31,0.41519) (32,0.41519) (33,0.41519) (34,0.41519) (35,0.41519) (36,0.41519) (37,0.41519) (38,0.41519) (39,0.41519) (40,0.41161) (41,0.41161) (42,0.41161) (43,0.41161) (44,0.41161) (45,0.41161) (46,0.41161) (47,0.41161) (48,0.41161) (49,0.41161) (50,0.41161) (51,0.41161) (52,0.41161) (53,0.41161) (54,0.41161) (55,0.41161) (56,0.41161) (57,0.41161) (58,0.34161) (59,0.34161) (60,0.47049) (61,0.47049) (62,0.47049) (63,0.47049) (64,0.47049) (65,0.47049) (66,0.47049) (67,0.47049) (68,0.47049) (69,0.47049) (70,0.47049) (71,0.47049) (72,0.47049) (73,0.47049) (74,0.47049) (75,0.47049) (76,0.47049) (77,0.47049) (78,0.47049) (79,0.47049) (80,0.47049) (81,0.47049) (82,0.47049) (83,0.47049) (84,0.47049) (85,0.47049) (86,0.47049) (87,0.47049) (88,0.41357) (89,0.46838) (90,0.46838) (91,0.46838) (92,0.46838) (93,0.46838) (94,0.46838) (95,0.46838) (96,0.46838) (97,0.46838) (98,0.46838) (99,0.46838) (100,0.46838) (101,0.46838) (102,0.46838) (103,0.46838) (104,0.46838) (105,0.46838) (106,0.46838) (107,0.46838) (108,0.46838) (109,0.46838) (110,0.46838) (111,0.46838) (112,0.46838) (113,0.46838) (114,0.46838) (115,0.46838) (116,0.43733) (117,0.42912) (118,0.42912) (119,0.42912) (120,0.42912) (121,0.42912) (122,0.42912) (123,0.42912) (124,0.42912) (125,0.42912) (126,0.42912) (127,0.42912) (128,0.42912) (129,0.42995) (130,0.42995) (131,0.42995) (132,0.42995) (133,0.42995) (134,0.42995) (135,0.42995) (136,0.42995) (137,0.45465) (138,0.45356) (139,0.45356) (140,0.45356) (141,0.45356) (142,0.45356) (143,0.45356) (144,0.45356) (145,0.45356) (146,0.45356) (147,0.45356) (148,0.45356) (149,0.45356) (150,0.45356) (151,0.45356) (152,0.45356) (153,0.45662) (154,0.45662) (155,0.45885) (156,0.45885) (157,0.45885) (158,0.45885) (159,0.45885) (160,0.45885) (161,0.45885) (162,0.45885) (163,0.45885) (164,0.45885) (165,0.45885) (166,0.45885) (167,0.45885) (168,0.45885) (169,0.45885) (170,0.48507) (171,0.48507) (172,0.48507) (173,0.48507) (174,0.48507) (175,0.48196) (176,0.48196) (177,0.48196) (178,0.48196) (179,0.48196) (180,0.48595) (181,0.48595) (182,0.48595) (183,0.48595) (184,0.48595) (185,0.48595) (186,0.48484) (187,0.48484) (188,0.48484) (189,0.48484) (190,0.48484) (191,0.48484) (192,0.48484) (193,0.48484) (194,0.48484) (195,0.48484) (196,0.48484) (197,0.48484) (198,0.48484) (199,0.48484) (200,0.46316) (201,0.47885) (202,0.47885) (203,0.47885) (204,0.47885) (205,0.47885) (206,0.46531) (207,0.46531) (208,0.46531) (209,0.46531) (210,0.46531) (211,0.46531) (212,0.44321) (213,0.44321) (214,0.44321) (215,0.44321) (216,0.4713) (217,0.4713) (218,0.4713) (219,0.4713) (220,0.4713) (221,0.46836) (222,0.46836) (223,0.46836) (224,0.46836) (225,0.46836) (226,0.46836) (227,0.46836) (228,0.47393) (229,0.47393) (230,0.47393) (231,0.47393) (232,0.47393) (233,0.47393) (234,0.47393) (235,0.47393) (236,0.47393) (237,0.47393) (238,0.4571) (239,0.47467) (240,0.46496) (241,0.46496) (242,0.46496) (243,0.46496) (244,0.46496) (245,0.46496) (246,0.46985) (247,0.46985) (248,0.46985) (249,0.46985) (250,0.46985) (251,0.47807) (252,0.47807) (253,0.47807) (254,0.47807) (255,0.47807) (256,0.47807) (257,0.47807) (258,0.47807) (259,0.47807) (260,0.47807) (261,0.47807) (262,0.47807) (263,0.47807) (264,0.47807) (265,0.47807) (266,0.46479) (267,0.46479) (268,0.46479) (269,0.46479) (270,0.46479) (271,0.46479) (272,0.46479) (273,0.46479) (274,0.46479) (275,0.46479) (276,0.46479) (277,0.46479) (278,0.46479) (279,0.46163) (280,0.45201) (281,0.45201) (282,0.45201) (283,0.45201) (284,0.45946) (285,0.45946) (286,0.45946) (287,0.45946) (288,0.45946) (289,0.45946) (290,0.45946) (291,0.46359) (292,0.46359) (293,0.45821) (294,0.45821) (295,0.45821) (296,0.45821) (297,0.45821) (298,0.45821) (299,0.45821)};
\legend{Proposed method,CCGA,CACD,GA-net}
\end{axis}
\end{tikzpicture}
\caption{Comparison of the results of the proposed method with other GAs on the Polbooks dataset. We ran all of the methods by population size of 100 for 300 generations.}
\label{FIG:11}
\end{figure*}
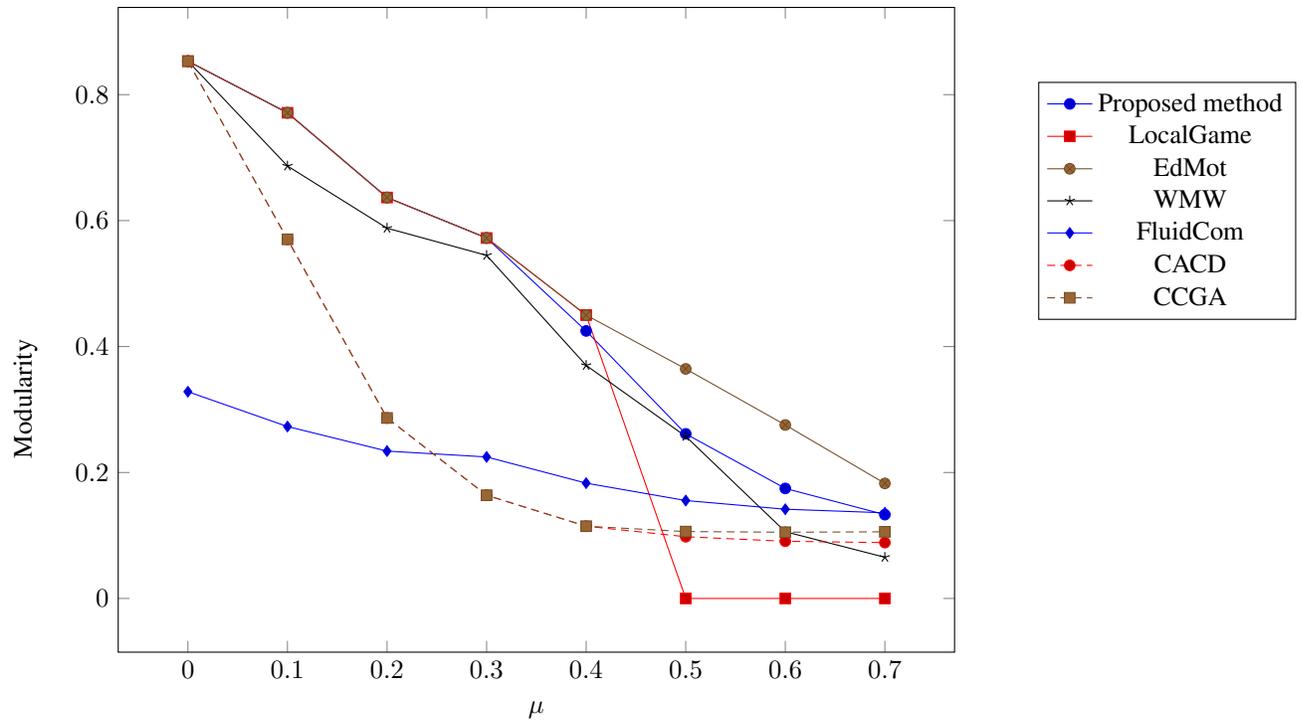
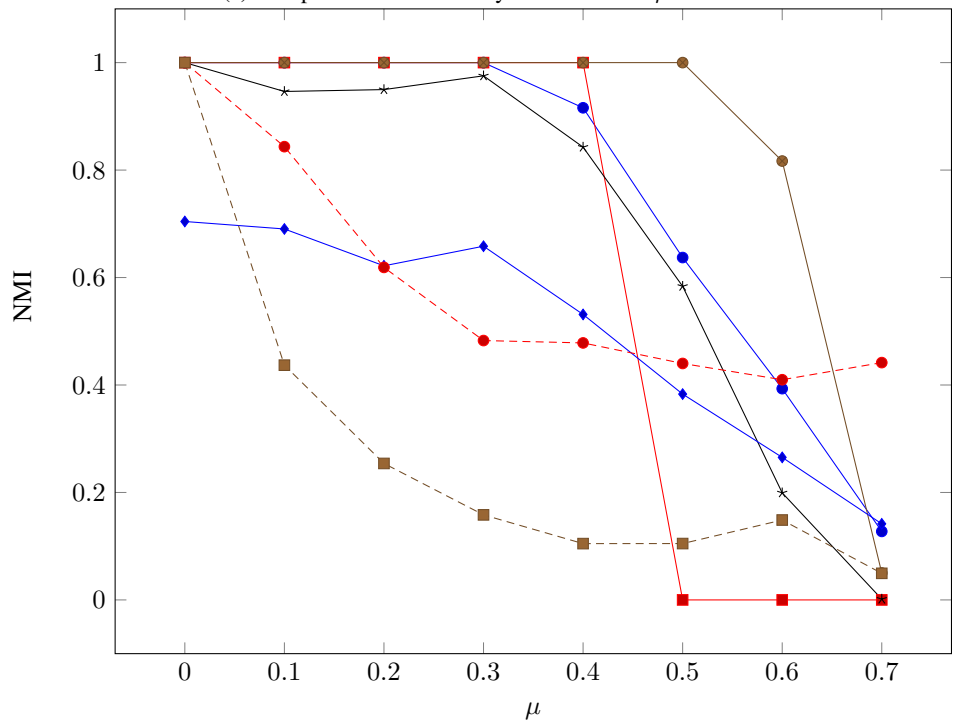
\begin{figure*}[t!]
\begin{subfigure}[t]{5in}
\begin{tikzpicture}[left]
\begin{axis}[
width=5in,
height=4in,
legend columns=1,
legend style={at={(1.1,0.70)},anchor=west},xlabel=$\mu$, ylabel=Modularity]
%GenMST
\addplot coordinates {
(0,0.8532) (0.1,0.7712) (0.2,0.6367) (0.3,0.5724) (0.4,0.4248) (0.5,0.2614) (0.6,0.1747) (0.7,0.1331)
};
%LocalGame
\addplot coordinates {
(0,0.8532) (0.1,0.7712) (0.2,0.6367) (0.3,0.5724) (0.4,0.45) (0.5,0.0) (0.6,0.0) (0.7,0.0)
};
%EdMot
\addplot coordinates {
(0,0.8532) (0.1,0.7712) (0.2,0.6367) (0.3,0.5724) (0.4,0.4500) (0.5,0.3645) (0.6,0.2755) (0.7,0.1827)
};
%WMW
\addplot coordinates {
(0,0.8532) (0.1,0.6867) (0.2,0.5880) (0.3,0.5448) (0.4,0.3701) (0.5,0.2574) (0.6,0.1060) (0.7,0.0654)
};
%FluidCom
\addplot coordinates {
(0,0.3283) (0.1,0.2730) (0.2,0.2340) (0.3,0.2247) (0.4,0.1831) (0.5,0.1555) (0.6,0.1417) (0.7,0.1360)
};
%CACD
\addplot coordinates {
(0,0.8532) (0.1,0.570256) (0.2,0.286598) (0.3,0.163895) (0.4,0.11474) (0.5,0.0980032) (0.6,0.0908123) (0.7,0.0886888)
};
% CCGA
\addplot coordinates {
(0,0.8532) (0.1,0.570256) (0.2,0.286598) (0.3,0.163895) (0.4,0.11474) (0.5,0.1062) (0.6,0.1052) (0.7,0.1058)
};
\legend{Proposed method ,LocalGame,EdMot,WMW,FluidCom,CACD,CCGA}
\end{axis}
\end{tikzpicture}
\subcaption{Comparison of modularity value for LFR-$\mu$ dataset}
\end{subfigure}
\\
\begin{subfigure}[t]{5in}
\begin{tikzpicture}[right]
\begin{axis}[
width=5in,
height=4in,
xlabel=$\mu$, ylabel=NMI]
%GenMST
\addplot coordinates {
(0,1.0) (0.1,1.0) (0.2,1.0) (0.3,1.0) (0.4,0.9159) (0.5,0.6372) (0.6,0.3932) (0.7,0.1275)
};
%LocalGame
\addplot coordinates {
(0,1.0) (0.1,1.0) (0.2,1.0) (0.3,1.0) (0.4,1.0) (0.5,0) (0.6,0) (0.7,0)
};
%EdMot
\addplot coordinates {
(0,1.0) (0.1,1.0) (0.2,1.0) (0.3,1.0) (0.4,1.0) (0.5,1.0) (0.6,0.8168) (0.7,0.0501)
};
%WMW
\addplot coordinates {
(0,1.0) (0.1,0.9463) (0.2,0.9498) (0.3,0.9754) (0.4,0.8427) (0.5,0.5837) (0.6,0.1993) (0.7,0.0015)
};
%FluidCom
\addplot coordinates {
(0,0.7043) (0.1,0.6904) (0.2,0.6218) (0.3,0.6583) (0.4,0.5313) (0.5,0.3831) (0.6,0.2653) (0.7,0.1416)
};
%CACD
\addplot coordinates {
(0,1.0) (0.1,0.843581) (0.2,0.618857) (0.3,0.482719) (0.4,0.478312) (0.5,0.440063) (0.6,0.409809) (0.7,0.441668)
};
% CCGA
\addplot coordinates {
(0,1.0) (0.1,0.4369) (0.2,0.2540) (0.3,0.1583) (0.4,0.1048) (0.5,0.1049) (0.6,0.1487) (0.7,0.0495)
};
\end{axis}
\end{tikzpicture}
\subcaption{Comparison of NMI value for LFR-$\mu$ dataset}
\end{subfigure}
\caption{Comparison of the results of the proposed method with some of the recent community detection algorithms based on their response to different $\mu$ values of LFR networks.}
	\label{FIG:12}
\end{figure*}
Table-\ref{tbl4} and -\ref{tbl5} provide the results of comparisons of the proposed method with some of the other methods, based on NMI and modularity measures. We have used the \emph{Jaccard} similarity to assign weights to the edges, and the \emph{sine-based} mutation function to perform mutations in our experiments. For running all of the GA-based methods, we set the population size to 100 for small networks and 300 for larger ones. We ran all of the GA-based algorithms to at most 300 generations several times and recorded the best outcome. In our method, we set $\delta=0.1$ for our experiments. It can be seen that our method can outperform all of the other GA-based methods in almost all of the experiments. The proposed method's merits get explicit as the network size increases. This is especially obvious on the LFR5 and LFR6 datasets, which have a very high number of inter-community edges. The reason for the other methods' poor performance in these datasets is that the \emph{locus-based} methods keep testing different neighbors, and therefore they need both larger populations and more generations to reach better results in such datasets.

Figure-\ref{FIG:11} shows another significant analysis of the different GA-based community detection algorithms. This figure shows the modularity of the best individual of each GA-based method on the Polbooks dataset over several generations. For this dataset, we set the generation size to 100 and let the methods run for 300 generations. Generation number 0 depicts the performance of initial population generation functions. While the best individual of our initial population has individuals with modularity 0.5 (almost equal to the final modularity), the closest method (CCAD) can produce individuals with at most 0.43 modularity value. On the other hand, while our method converges at 20-th generation with 0.5234, CCAD can only beat our method on 148-th generation, reaching convergence at 189-th generation on the modularity of 0.5266, the next two algorithms can never find better partitions even on 300-th generation. Therefore, comparisons prove that the proposed algorithm can produce better results compared with other GAs. It worth mentioning that the GA-net method's high fluctuations are the result of its fitness function. As we explained in Section-\ref{secRW}, this method doesn't use modularity as its fitness function (CACD and CCGA are both \emph{modularity-optimization-based} methods).

Comparing our method with other state-of-the-art methods shows that the proposed method can produce competitive results on both NMI and modularity measures. On the real-world datasets, our method results in an average of 0.607 modularity, while, the best modularity is produced by the Louvain algorithm, which reaches 0.617. But, while our method's average NMI value is 0.634, Louvain results in 0.618. The best NMI value belongs to CACD with 0.826 NMI and 0.56 modularity. But, on the synthetic datasets, our method can reach the highest modularity of 0.62 while lagging behind the best NMI value of 0.97 with 0.92. Therefore our method can be considered as a successful GA-based community detection algorithm.

In synthetic networks, $\mu$ defines the ratio of the inter-community edges. In Figure-\ref{FIG:10}, we compare the effect of the $\mu$ parameter on the quality of the outcomes of our method. As it can be seen from the figure, an increase in the $\mu$ decreases modularity on almost all of the methods. This is due to the fact that more inter-community edges are the exact opposite of the main definition of the community. LocalGame, a \emph{similarity-based} method because of its high dependency on the distance between the communities, falls immediately to 0 on both measures as $\mu$ goes beyond 0.5. CCGA and CACD behave almost identically to the $\mu$ parameter on modularity, but CACD results in better NMI values. Yet, even though CACD's NMI value stays the highest at $\mu=0.7$, but its NMI and modularity values start to decrease so quickly on the $\mu=0.1$. The proposed method's NMI and modularity remain the highest until $\mu$ is below 0.5, and after passing 0.5 it starts to dwindle, yet staying as the second-best method for almost all of the other $\mu$ values. Comparing the results shows that the proposed algorithm can result in acceptable outcomes on both measures.

Extensive comparisons on both real-world and synthetic datasets show that the proposed method can produce competitive results with state-of-the-art algorithms. In the next section, we perform some statistical tests on these results.

\subsection{Statistical tests}
\label{STAT}

After presenting the results of the comparisons in \ref{COMP}, here, we are going to show the results of some statistical tests on the results. We performed multiple comparisons on the results of the algorithms. Multiple comparisons can be performed on different methods \cite{scmamp}. Here, we have used the Friedman test, along with the post-hoc Nemenyi test. We conducted the tests separately on NMI and Modularity results for both real-world and synthetic datasets.

The Friedman test is a non-parametric statistical test that has been introduced to conclude whether any significant difference among different treatments (algorithms in our case) exists or not (at least two of the treatments are significantly different or not). We have used the Friedman test with Iman \& Davenport extension, which is known to be an omnibus test. Table-\ref{tbl6} shows the results of the Friedman test on the data presented in \ref{COMP} with $\alpha = 0.05$. Since all of the \emph{p-values} are smaller than $0.05$, we can say that the Friedman test has confirmed a significant difference between at least two algorithms in all cases.

\begin{table}%[cols=5,pos=h]%,cols=3,pos=h
\centering
\caption{The results of the Friedman test on the NMI and Modularity outcomes of the algorithms. All of the tests show the existence of a significant difference between at least two algorithms.}\label{tbl6}
\begin{tabular}{c c c c c} 
\hline
Test & real-world (NMI)& real-world (Q)& synthetic (NMI)& synthetic (Q)\\
\hline
\emph{p-value} & $0.0007059$ &$1.208e-09$ & $0.0004695$ & $0.0007831$  \\
\hline
\end{tabular}
\end{table}

\begin{figure*}[t!]
\centering
  \includegraphics[width=5in,height=3in]{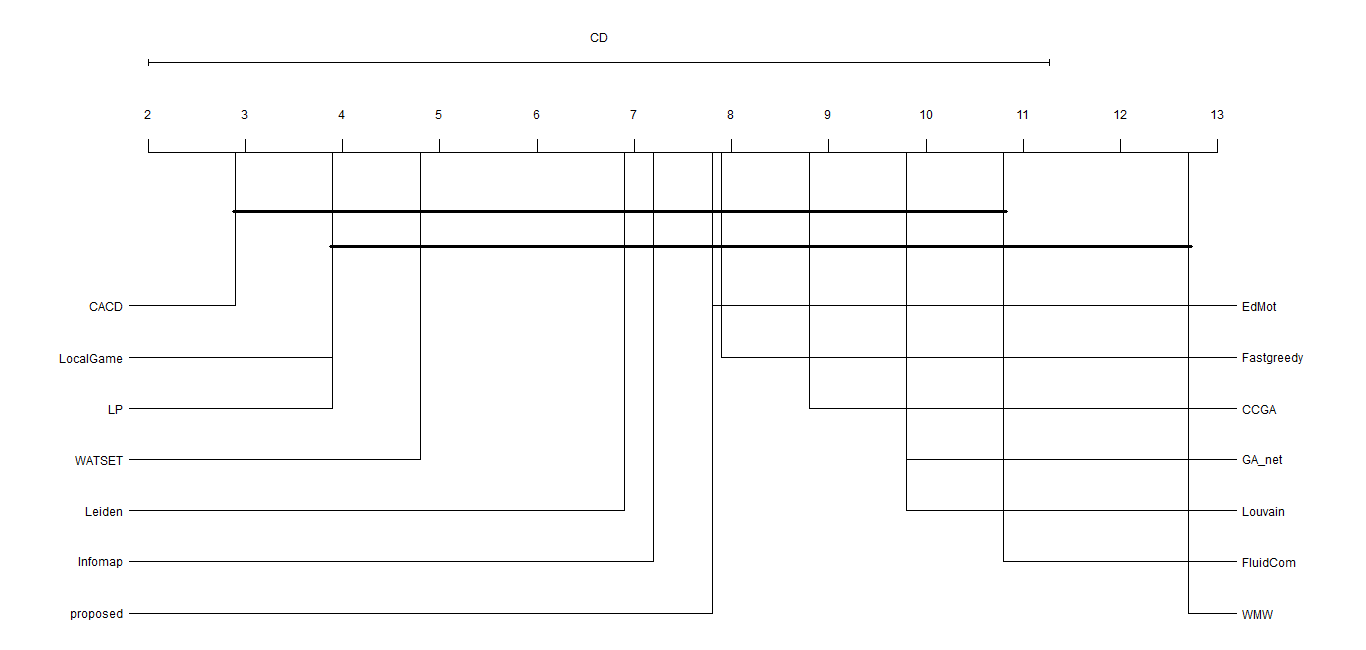}
\caption{Critical distance diagram for the Nemneyi test on NMI outcomes of the algorithms on real-world datasets. The CD value is $9.273$}
\label{FIG:13}
\end{figure*}

\begin{figure*}[t!]
\centering
  \includegraphics[width=5in,height=3in]{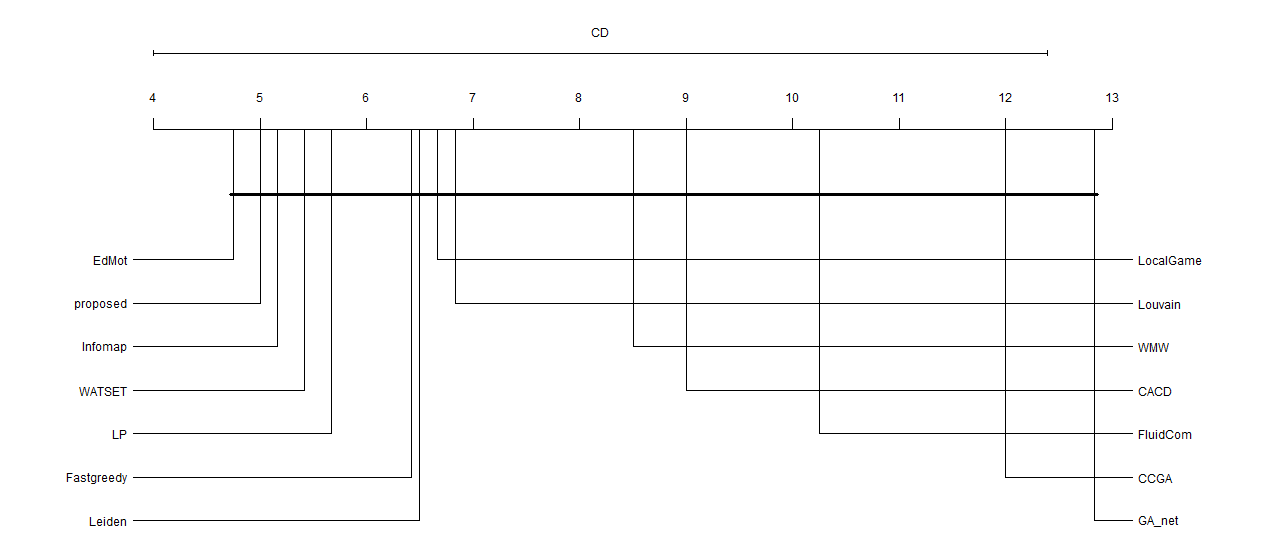}
\caption{Critical distance diagram for the Nemneyi test on Modularity (Q) outcomes of the algorithms on synthetic datasets. The CD value is $8.39$}
\label{FIG:16}
\end{figure*}

Despite being a powerful statistical test, the Friedman test can not exactly show which treatments differ from each other. For this reason, we have conducted two post-hoc tests. One of the most famous post-hoc tests is the Nemenyi test. The Nemenyi computes a distance value for each pair of treatments and a critical distance (CD) value. Treatments with a higher distance than the critical distance value are considered \emph{significantly different} from each other. We have performed the test on NMI and Modularity outcomes of the algorithms on all datasets. Figure-\ref{FIG:13} to Figure-\ref{FIG:16} show the critical distance diagram of the tests. 

\begin{figure*}[t!]
\centering
  \includegraphics[width=5in,height=3in]{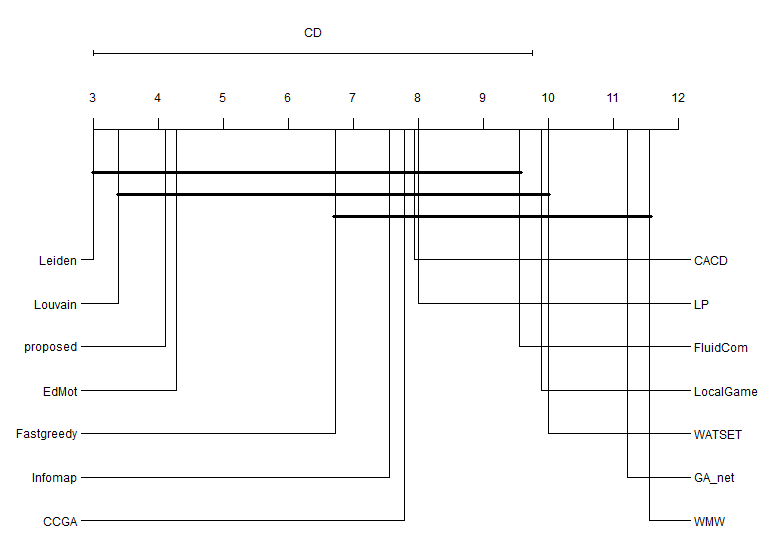}
\caption{Critical distance diagram for the Nemneyi test on Modularity (Q) outcomes of the algorithms on real-world datasets. The CD value is $6.7615$}
\label{FIG:14}
\end{figure*}

\begin{figure*}[t!]
\centering
  \includegraphics[width=5in,height=3in]{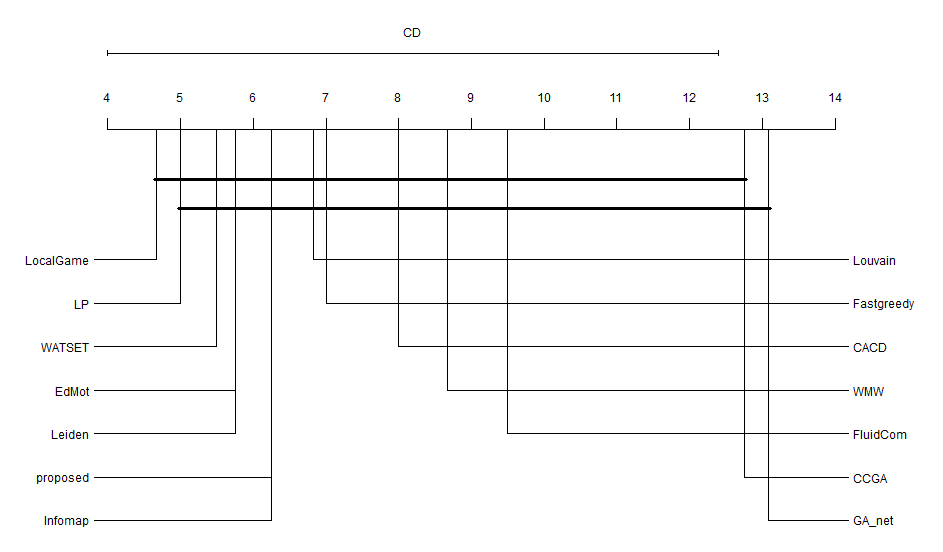}
\caption{Critical distance diagram for the Nemneyi test on NMI outcomes of the algorithms on synthetic datasets. The CD value is $8.39$}
\label{FIG:15}
\end{figure*}

The critical distance value for the Nemenyi test on NMI values of the real-world datasets is $9.273$, and the only two algorithms with a significant difference are \emph{CACD and WMW} with a distance of $9.8$ (Figure-\ref{FIG:13}). The outcome was predictable considering that CACD had an average NMI value of $0.826$ while WMW had achieved an average of $0.526$. On the other hand, Figure-\ref{FIG:14} shows the critical distance diagram of the algorithms based on their Modularity results. Here, we can see that some algorithms have outperformed the others significantly. Our method has shown to have a significant difference with \emph{GA-net} and \emph{WMW} resulting in a difference of $7.11$ and $7.44$ while the critical distance value is $6.76$. A significant difference has also been detected between \emph{EdMot and GA-Net},  \emph{Leiden and GA-Net}, \emph{Louvain and GA-Net}, \emph{Leiden and WATSET}, \emph{EdMot and WMW}, \emph{Leiden and WMW}, \emph{Louvain and WMW}, and \emph{Leiden and LocalGame}. The critical difference diagram of the algorithms on their NMI values on synthetic datasets is shown in Figure-\ref{FIG:15}. Here, the only pair of algorithms that has shown a significant distance from each other is the pair of \emph{LocalGame and GA-net} that has reached a distance of $8.4167$ while the critical distance is $8.3912$. Figure-\ref{FIG:16} shows the critical distance diagram of the algorithms on the Modularities of the synthetic datasets. Here, the Nemenyi test hasn't detected any algorithm to outperform the others. The highest difference is between \emph{EdMot and GA-Net} with $8.083$ while our method's distance from \emph{GA-Net} is $7.83$.

Analyzing the results and the statistical tests show that our method is comparable with the state-of-the-art methods, and it is capable of showing significantly better outcomes (considering the average NMI and Modularity values in Table-\ref{tbl4} and Table-\ref{tbl5}) compared with some of the other methods, especially some genetic-based ones. Some of the best-known methods, which have been designed to maximize the Modularity, such as Louvain, Leiden, and EdMot, haven't been able to show significantly different outcomes compared with the presented methods. In the next section, we will conclude the paper.

\section{Conclusions}
\label{CONC}
In this paper, we proposed a new method to detect communities in complex networks using an adaptive evolutionary approach. Here, we have introduced a new encoding scheme, using node similarity measures and MSTs, which does not have deficiencies such as resulting in the separate communities and meaningless mutations that were likely to happen in \emph{locus-based} and \emph{solution-vector-based} representations. Also, we have introduced a new method to generate the initial population. This method can enhance the convergence time and quality of the initial population drastically. Furthermore, an adaptive \emph{sine-based} mutation function was introduced, which changes the mutation probability of each gene based on the depth of its corresponding edge in MST. The new mutation function can help to reach better and faster results.
Several experiments were conducted on the real-world and synthetic networks, and results were compared with other community detection algorithms. Results show that the proposed method can perform better and faster than the existing GA-based methods and produce comparable results with the state-of-the-art methods on both NMI and modularity measures.

GAs usually consist of several stages such as initial population generation, selection, crossover, and mutation. The ability to use different combinations of functions in each stage gives GAs advantageous flexibility. Therefore as a future work of the proposed method, we consider an in-detail analysis of the effects of different functions per each stage. 
Evaluation of the impact of different similarity measures along with the extraction of application-specific skeleton networks, instead of MSTs, to be used in different types of networks (such as temporal networks, directed networks, etc.) are among other possible future works of this paper.

\bibliographystyle{model1-num-names}
\bibliography{paper.bib}

\end{document}